%% file: main.tex
\let\TeXyear\year
\documentclass{ieeeaccess}
\let\setyear\year
\let\year\TeXyear

\usepackage{amsmath,amsthm,amssymb,amsfonts}   
\usepackage{bm}                         
\usepackage{graphicx}                   
\usepackage{psfrag}                     
\usepackage{bbm}						
\usepackage[font=small]{caption}
\usepackage{subcaption}
\usepackage{hyperref}
\usepackage{todonotes}
\usepackage{url}
\usepackage{placeins}
\usepackage{textcomp}
\usepackage{siunitx}
\usepackage[stretch=20,shrink=20]{microtype}
\usepackage{tikz}
\usepackage{ifthen}
\usetikzlibrary{patterns}
\usetikzlibrary{shapes, arrows, chains, fit, positioning, calc, decorations.pathreplacing}
\usepackage{soul}
\usepackage{comment}
\usepackage{lipsum}
\usepackage{cite} 
\usepackage[nameinlink]{cleveref}
\usepackage[normalem]{ulem}
\usepackage{empheq}
\definecolor{accessblue}{cmyk}{1, 0.3, 0, 0.2}
\definecolor{greycolor}{cmyk}{0,0,0,.8}

\makeatletter
\AtBeginDocument{\DeclareMathVersion{bold}
\SetSymbolFont{operators}{bold}{T1}{times}{b}{n}
\SetSymbolFont{NewLetters}{bold}{T1}{times}{b}{it}
\SetMathAlphabet{\mathrm}{bold}{T1}{times}{b}{n}
\SetMathAlphabet{\mathit}{bold}{T1}{times}{b}{it}
\SetMathAlphabet{\mathbf}{bold}{T1}{times}{b}{n}
\SetMathAlphabet{\mathtt}{bold}{OT1}{pcr}{b}{n}
\SetSymbolFont{symbols}{bold}{OMS}{cmsy}{b}{n}
\renewcommand\boldmath{\@nomath\boldmath\mathversion{bold}}}
\makeatother

\def\BibTeX{{\rm B\kern-.05em{\sc i\kern-.025em b}\kern-.08em
    T\kern-.1667em\lower.7ex\hbox{E}\kern-.125emX}}

\newcommand{\compl}{\mathbb{C}}         
\newcommand{\real}{\mathbb{R}}          
\newcommand{\trans}{^{\text{T}}}		
\newcommand{\herm}{^{\text{H}}}			
\newcommand{\diag}[1]{\text{diag}\{#1\}}
\newcommand{\ftot}{\bm{\psi}_\text{t}}		
\newcommand{\finc}{\bm{\psi}_\text{i}}		
\newcommand{\fsc}{\bm{\psi}_\text{s}}		
\newcommand{\Groi}{\bm{G}_\Gamma^\Gamma} 
\newcommand{\Garray}{\bm{G}_\Gamma^\text{A}} 

\newcommand{\blue}[1]{#1}

\definecolor{myyellow}{rgb}{1, 1, .0}

\DeclareMathOperator*{\argmin}{argmin}
\DeclareMathOperator*{\argmax}{argmax}

\theoremstyle{definition}
\newtheorem{definition}{Definition}[section]
\newtheorem{theorem}{Theorem}

\begin{document}

\history{Date of publication xxxx 00, 0000, date of current version xxxx 00, 0000.}
\doi{10.1109/ACCESS.2024.3471430}
\setyear{2024}
\title{Misspecification of Multiple Scattering in Scalar Wave Fields and its Impact in Ultrasound Tomography}
\author{
\uppercase{Eduardo P\'{e}rez}\authorrefmark{1,2}, \IEEEmembership{Student Member, IEEE}, \uppercase{Sebastian Semper}\authorrefmark{2},
\uppercase{Sayako Kodera}\authorrefmark{1}, \uppercase{Florian R\"{o}mer}\authorrefmark{1}, \IEEEmembership{Senior Member, IEEE},
\uppercase{Giovanni Del Galdo}\authorrefmark{2}, \IEEEmembership{Member, IEEE}}

\address[1]{Fraunhofer Institute for Nondestructive Testing IZFP, Saarbr\"{u}cken, Germany}
\address[2]{Technische Universit\"{a}t Ilmenau, Ilmenau, Germany}
\tfootnote{This work was supported by the Fraunhofer Internal Programs under the grant Attract 025-601128 and by the Thuringian Ministry of Economic Affairs, Science and Digital Society (TMWWDG). The work of Sebastian Semper was supported by DFG through the project ``JCRS CoMP'' under grant TH 494/35-1. \iffalse{We acknowledge support for the Article Processing Charge by the Open Access Publication Fund of the Technische Universit\"{a}t Ilmenau.}\fi}

\markboth
{E. P\'{e}rez \headeretal: Misspecification of Multiple Scattering in Scalar Wave Fields and its Impact in Ultrasound Tomography}
{E. P\'{e}rez \headeretal: Misspecification of Multiple Scattering in Scalar Wave Fields and its Impact in Ultrasound Tomography}

\corresp{Corresponding author: Eduardo P\'{e}rez (e-mail: eduardo.perez@tu-ilmenau.de).}

\input{sections/abstract}
\titlepgskip=-21pt

\maketitle%


\input{sections/introduction}
\input{sections/MCRB_background}
\input{sections/models}
\input{sections/MCRB}
\input{sections/simulations}
\input{sections/discussion}

\input{sections/conclusion}


\bibliographystyle{IEEEtran}
\bibliography{mybibfile}

\begin{IEEEbiography}%
	[{\includegraphics[width=1in,height=1.25in,
        trim={{4cm} {5cm} {4cm} {2.5cm}},
        clip,keepaspectratio]{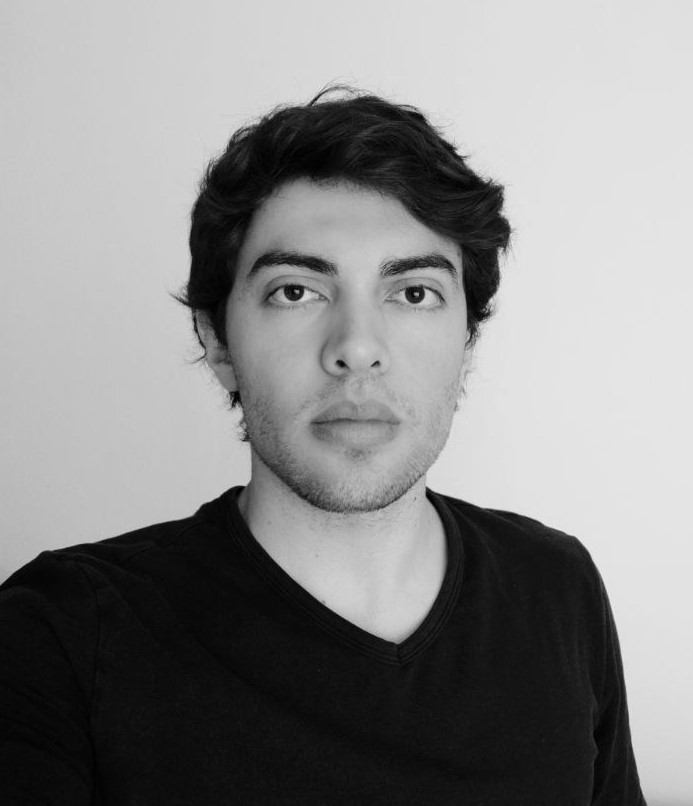}}]%
        {Eduardo~Pérez}
(S’19) received the M.Sc.~degree in communications~and~signal~processing at the Technische Universität Ilmenau, Ilmenau, Germany, in 2019, where he currently pursues a doctoral degree in signal~processing. 

He joined the SigMaSense group at the Fraunhofer Institute for Nondestructive Testing IZFP and the Electronic~Measurements~and~Signal~Processing group at TU~Ilmenau in 2020. His research interests include array signal processing, compressed sensing, parameter estimation and machine learning.
\end{IEEEbiography}

\begin{IEEEbiography}%
  [{\includegraphics[width=1in,height=1.25in,
  trim={{0} {0} {0} {0}},
  clip,keepaspectratio]{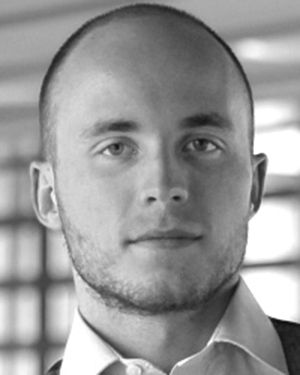}}]%
        {Sebastian~Semper}
studied mathematics at Technische Universit\"at Ilmenau, Ilmenau, Germany. There, he received the Master of Science degree in 2015. In 2022 he finished his doctoral studies and received the doctoral degree with honors in electrical engineering. Since then, he has been a post doctoral student in the Electronic Measurements and Signal Processing (EMS) Group, a joint research activity between the Fraunhofer Institute for Integrated Circuits IIS and TU Ilmenau.

Since 2015, he has been a Research Assistant with the EMS. His research interests consist of compressive sensing, parameter estimation, optimization, numerical methods and algorithm design.
\end{IEEEbiography}

\begin{IEEEbiography}%
        [{\includegraphics[width=1in,height=1.25in,
        trim={{2.5cm} {1cm} {3.6cm} {0.5cm}},
        clip,keepaspectratio]{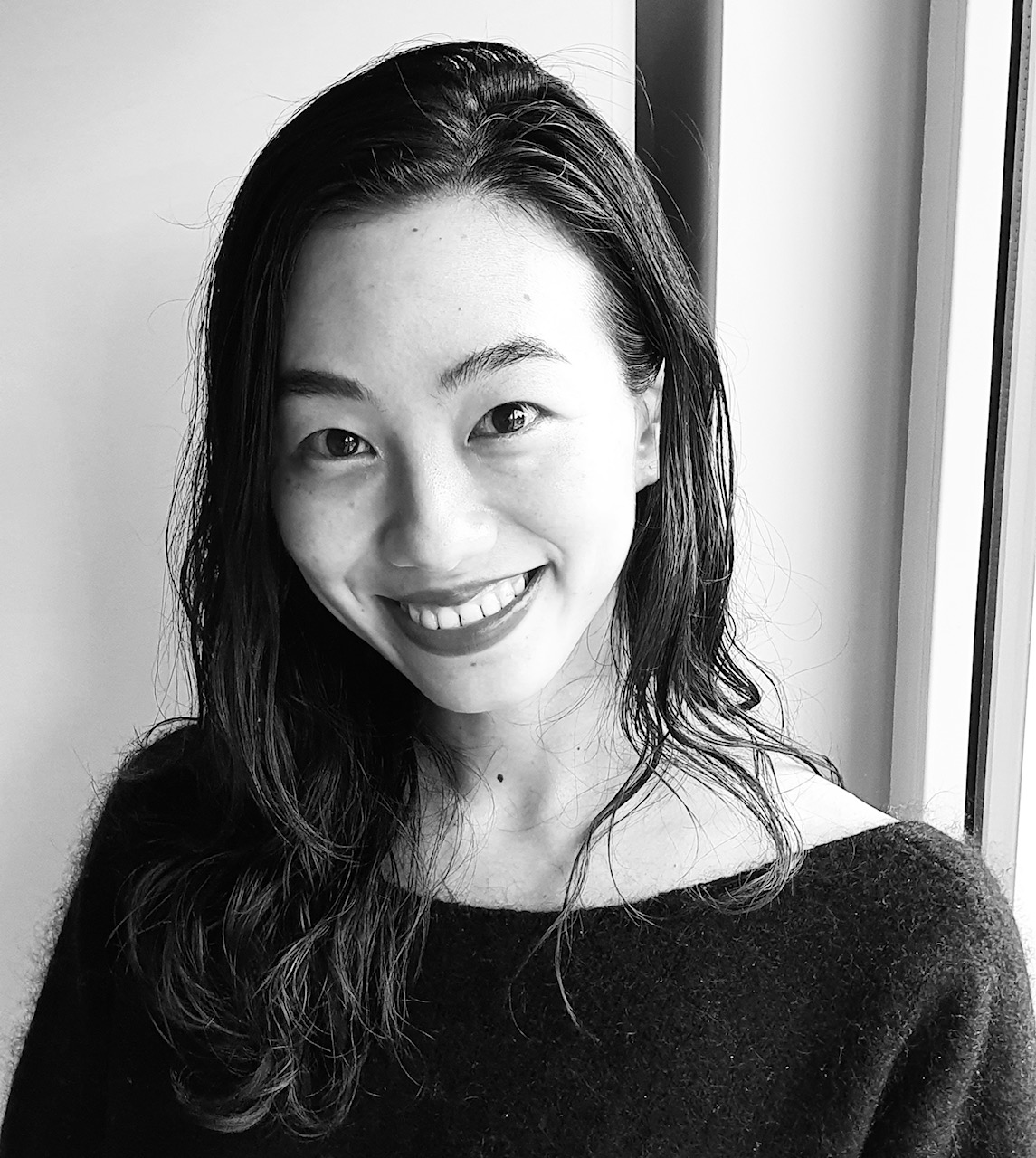}}]%
        {Sayako~Kodera}
received the M.Sc. degree in communications and signal processing at the Technische Universit\"{a}t Ilmenau, Ilmenau, Germany, in 2021. 

In September 2021, she joined the SigMaSense group at the Fraunhofer Institute for Nondestructive Testing IZFP and the Electronic Measurements and Signal Processing group at TU Ilmenau. Her research interests include statistical signal processing, robust optimization, information geometry and model-based deep learning.
\end{IEEEbiography}

\begin{IEEEbiography}%
  [{\includegraphics[width=1in,height=1.25in,clip,keepaspectratio]{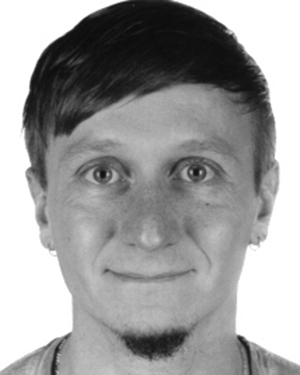}}]%
        {Florian~Römer}
(S'04 - M'13 - SM'16) received the doctoral (Dr.-Ing.) degree in electrical engineering from Technische Universit\"{a}t Ilmenau in 2012. In 2012 he joined the Electronic Measurements and Signal Processing Group, a joint research activity between the Fraunhofer Institute for Integrated Circuits IIS and TU Ilmenau, as a postdoctoral research fellow. 

From 2006 to 2012, he was a Research Assistant in the Communications Research Laboratory at TU Ilmenau. In 2018 he joined the Fraunhofer Institute for Nondestructive Testing IZFP where he leads the SigMaSense group with a research focus on innovative sensing and signal processing for nondestructive testing and has recently taken the role of Chief Scientist leading the center of expertise on applied AI, signal processing and data analysis. 

Mr. R\"{o}mer has served as associate editor for IEEE Transactions on Signal Processing from 2020-2022 and as senior area editor from 2022.
\end{IEEEbiography}

\begin{IEEEbiography}%
  [{\includegraphics[width=1in,height=1.25in,clip,keepaspectratio]{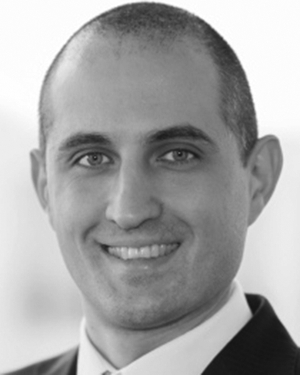}}]
        {Giovanni~Del~Galdo}
(M’12) received the Laurea degree in telecommunications engineering from the Politecnico di Milano, Milan, Italy, in 2002, and the doctoral degree in MIMO channel modeling for mobile communications from Technische Universität Ilmenau, Ilmenau, Germany, in 2007. 

He then joined the Fraunhofer Institute for Integrated Circuits, Erlangen, Germany, focusing on audio watermarking and parametric representations of spatial sound. Since 2012, he leads a joint research group composed of a Department at Fraunhofer Institute for Integrated Circuits IIS and, as a Full Professor, a Chair with TU Ilmenau on the research area of electronic measurements and signal processing. His research interests include the analysis of multidimensional signals, over-the-air testing for terrestrial and satellite communication systems, and sparsity promoting reconstruction methods.
\end{IEEEbiography}
\EOD
\end{document}

%% file: sections/abstract.tex
\begin{abstract}
{In this work, we investigate the localization of point-like targets in the presence of multiple scattering.} We focus on the often omitted scenario in which measurement data is affected by multiple scattering, and a simpler model is employed in the estimation. We study the impact of such model mismatch by means of the Misspecified \emph{Cram\'{e}r-Rao Bound} (MCRB). {In numerical simulations inspired by tomographic inspection in Ultrasound Nondestructive Testing (UNDT), the MCRB is shown to correctly describe the estimation variance of localization parameters under misspecification of the wave propagation model. We provide extensive discussion on the utility of the MCRB in the practical task of verifying whether a chosen misspecified model is suitable for localization based on the properties of the maximum likelihood estimator and the nuanced distinction between bias and parameter space differences. Finally, we highlight that careful interpretation is needed whenever employing the classical CRB in the presence of mismatch through numerical examples based on the Born approximation and other simplified propagation models stemming from it. {Although motivated by UNDT, the analysis and discussion throughout this work generally apply to localization tasks based on scalar wave field measurements.}}
\end{abstract}



%% file: sections/introduction.tex
\section{Introduction} \label{section_introduction}
In order to accurately post-process ultrasound measurement data, it is necessary to consider the intricate effects that arise {during propagation}. Such effects depend on physical parameters of the corresponding measurement scenario. Characteristic examples that have drawn the attention of researchers include shadowing, wherein a large defect obscures the presence of another \cite{shadowing, shadowing_jan}; reverberation, in which multiple scattering produces repeated echoes that clutter the data and make it difficult to interpret \cite{shadowing, reverberation}; and attenuation and dispersion, whose impact varies in severity based on the medium characteristics such as grain size in metals \cite{atten_tissue, metals1, metals2}.

Among the {signal processing} community, the phenomenon of multiple scattering has been of particular interest. On one hand, multiple scattering is seen as one of the root causes of image clutter and its negative effects must be alleviated in post-processing \cite{reverberation, reverberation_medical, reverberation_ndt}. On the other hand, some authors posit that the presence of multiple scattering can be beneficial, enabling more accurate estimates of the parameters of interest \cite{crb_mono1, crb_mono2}, and going as far as to propose measurement systems in which additional scatterers are introduced artificially in order to promote multiple scattering \cite{crb_resolved}. In both cases, practical data models and estimators are chosen with simplicity and computational efficiency in mind. In cases where multiple scattering is seen negatively, these practical models and estimators are applied to measurement data that follows a different model than the assumed one. On the contrary, when theoretical lower bounds are derived, it is often under the crucial assumption that the chosen data model perfectly matches the true model from which the data was drawn. 

More concisely: the two views are at odds due to model mismatch that is often ignored when deriving theoretical performance results. Practical data models are often implicitly or explicitly based on the first order Born approximation, under which multiple scattering is ignored \cite{simulation}. This simplifying assumption results in models and estimation algorithms that are simple and tractable, but it is only valid in limited regimes \cite{iterative_born, Pade}. {In practice, these models and estimators are applied to measurement data in order to formulate and solve an inverse problem that yields an image and/or estimates of physical parameters. As the data does not exactly match the model, the interpretability of the images and parameters estimates is negatively affected \cite{CURE}. The applicability of theoretical performance results in these practical, mismatched scenarios depends on whether the mismatch is included in the analysis.}

The works \cite{crb_optics, crb_mono1, crb_mono2, crb_resolved, born_crb} present results on the impact of multiple scattering on the estimation of parameters from scattered wave fields. The authors rely on the \emph{Cram\'{e}r-Rao Bound} (CRB) as their tool of choice, as it provides a lower bound on the variance that any unbiased estimator can achieve, and its computation is based only on the data model \cite{stoica}. Theoretical performance bounds of this kind are valuable in experiment design and benchmarking of common systems involving parameter estimation \cite{CRBsector, precision, CRBmimo, CRBcamera, CRBmdpi}. However, the usage of the classical CRB comes with the assumption that the data model used in the estimation task perfectly matches the true model underlying the data. In practice, this assumption translates into the two following cases. If multiple scattering is considered in the CRB, it is present both in the observed data and the fitted model. If multiple scattering is not considered, it is absent from both. These cases are mutually exclusive, meaning a direct comparison between the two may not provide exploitable knowledge. In addition, neither case considers the more practical scenario in which a model without multiple scattering is chosen for post-processing when multiple scattering is present in the data, in which case the model is \emph{misspecified}.

One way to alleviate the gap between theoretical performance bounds and the difficulties found when using practical estimators is to entirely replace the models and estimators with ones that are fully physically motivated, removing the mismatch. The field of seismology pioneered this approach through their \emph{Full Waveform Inversion} (FWI) techniques in which the estimators directly involve the wave equation \cite{seismic_overview, seismic}. More recently, medical \cite{fwi_original, fwi_new, bone, brain} and \emph{Nondestructive Testing} (NDT) \cite{guidedFWI, identificationFWI} ultrasound tomography techniques based on the same principle have been proposed. 

Although the FWI framework has the potential to greatly improve imaging quality, it comes at the cost of great computational complexity. FWI cannot be performed analytically, instead relying on iterative algorithms. Accurate estimates are obtained based on approximations to the deviation between the physical model and the observed data. This is necessary due to the complex, nonlinear nature of the model and, consequently, the highly non-convex nature of the target function being optimized. Furthermore, the forward model is computationally demanding. As a result, proper hardware infrastructure is required and reconstruction is time-consuming. What about cases in which such computational effort is inadmissible, e.g. when reconstructions must be done in near real-time or data volumes are overwhelmingly large, or scenarios in which the practice hints at simpler models yielding sufficiently good estimation results? In this work, {we employ the} \emph{Misspecified Cram\'{e}r-Rao Bound} (MCRB) as a tool to analyze the misspecification of multiple scattering.

The MCRB is an extension to the CRB that explicitly accounts for model misspecification. The framework for the MCRB was established more than $50$ years ago. However, it is only recently that it has drawn more interest thanks to {a wave of landmark publications regarding the MCRB and its usage \cite{richmond2013parameter, MCRBreal, MCRB, MCRBcomplex, mennad2018slepian,richmond2018constraints, gSlepian}}. Therein, the authors present the same key motivation behind the usage of the MCRB: there are scenarios where, even though the true model may be available, using it directly is prohibitively expensive. {This has sparked a variety of works investigating the impact of model misspecification, e.g. in time delay \cite{florianmcrb} and frequency \cite{harmonic} estimation.} Sharing this interest, we turn our focus to the scenario where a misspecified model is chosen on the grounds of being tractable, with the goal of studying the impact of this choice on the achievable performance. 

{The goal of this work is to study and discuss the impact of the misspecification of multiple scattering on the achievable point scatterer localization performance through the lens of the MCRB, and to do so in a manner that is accessible to the ultrasound NDT and signal processing communities.} In the interest of self-containedness, we review key elements of the MCRB and the modeling of ultrasound data. {However, the analysis presented in this work is transferrable to other localization tasks based on scalar wave fields, e.g. medical ultrasound imaging and radar localization.}

{We employ a generalized Slepian formula \cite{richmond2013parameter, gSlepian, mennad2018slepian}} for the MCRB of misspecified-unbiased estimators. We focus on the estimation scenarios where the assumed model has a misspecified mean $\bm{\mu}: \real^d \to \compl^N$ which differs from the true mean $\bm{s} \in \compl^N$, and both are afflicted by {circularly symmetric complex Gaussian noise with a known covariance matrix $\bm{\Sigma} \in \compl^{N \times N}$.} In particular, the true data model follows the scalar Helmholtz equation, while the misspecified model is the classical time delay model based on the first order Born approximation, consisting of scaled and shifted copies of a predetermined pulse shape. We then perform numerical experiments for an ultrasound NDT computed tomography scenario {consisting of two point-like scatterers, noting that the results obtained are valid for any number of scatterers so long as they are only pairwise close to each other}. Finally, we compare the MCRB to the classical CRB and discuss {the implications of the true and misspecified models having different parameter spaces, limiting the usefulness and interpretability of the classical CRB.}

%% file: sections/MCRB_background.tex
\section{Misspecified Cram\'{e}r-Rao Bound} \label{section_mcrbdef}
We begin by briefly discussing the MCRB and establishing the key components needed for its computation. For in-depth \blue{derivations of the MCRB and discussions on the class of estimators to which it applies, we refer the reader to \cite{vuong1986cramer, richmond2013parameter, MCRBreal, MCRB, MCRBcomplex, mennad2018slepian,richmond2018constraints, gSlepian} and the references therein}. The definitions and formulations presented next are based on the aforementioned references.

\subsection{General Definitions} \label{subsec_general_defs}
Denote by $\bm{x} \in \compl^{N}$ a single observation obeying {an unknown} \emph{probability density function} (PDF) $p_{X}(\bm{x})$, shortened as $p_{X}$ and referred to as the \emph{true PDF}. Instead of this true PDF, the observation $\bm{x}$ is \emph{assumed} to follow a PDF $f_X(\bm{x}|\bm{\theta})$, shortened as $f_{X \vert \bm{\theta}}$, which depends on the parameter vector $\bm{\theta} \in \Theta \subset \real^d$. {Throughout this work, we refer to the PDFs as \emph{statistical models}.} If there is no $\bm{\theta}$ for which $p_{X}(\bm{x}) = f_X(\bm{x}|\bm{\theta})$, the {assumed} model is said to be \emph{misspecified}. These conditions set the context for the following definitions and results.

\begin{definition}[\blue{Pseudo-True Parameter}] \label{def_PTS}
Consider the \emph{Kullback-Leibler Divergence} (KLD) from $f_{X \vert \bm{\theta}}$ to $p_X$ defined as
\begin{align}
	{D(p_X \Vert f_{X \vert \bm{\theta}}) \overset{\scriptscriptstyle\Delta}{=} \int \ln \left( \frac{p_X(\bm{x})}{f_X(\bm{x}\vert \bm{\theta})} \right) p_X(\bm{x}) \, d\bm{x}.}
\end{align}
If there exists a unique parameter $\bm{\theta}_0 \in \real^d$ such that 
\begin{align}
	\bm{\theta}_0 = \argmin_{\bm{\theta} \in \Theta} D(p_X \Vert f_{X \vert \bm{\theta}}),
\end{align}
then $\bm{\theta}_0$ is referred to as the \blue{\emph{Pseudo-True Parameter} (PTP).}
\end{definition}

The \blue{PTP} in \Cref{def_PTS} can be interpreted as the parameter of the misspecified model that best explains data stemming from the true model. We are interested in estimators that relate to the \blue{PTP} in the following way.

\begin{definition}[Misspecified Unbiasedness] \label{def_MU}
An estimator $\hat{\bm{\theta}}$ of $\bm{\theta}_0$ based on the misspecified model $f_{X \vert \bm{\theta}}$ is said to be unbiased in the misspecified case, or \emph{Misspecified Unbiased} (MS-unbiased), {if and only if}
\begin{align}
	\mathbb{E}_p \{ \hat{\bm{\theta}} \} = \int \hat{\bm{\theta}}(\bm{x}) p_X(\bm{x}) \, d\bm{x} = \bm{\theta}_0,
\end{align}
where $\mathbb{E}_p$ represents expectation with respect to the true PDF $p_X$. 
\end{definition}

The goal is to obtain a lower bound for the error covariance of MS-unbiased estimators following \Cref{def_MU}. To this end, the following auxiliary quantities are necessary.

\begin{definition}[Misspecified Log-Likelihood] \label{def_MLL}
The \emph{Misspecified Log-Likelihood} (MLL) $v:\real^d \to \real$ is given by
\begin{align}
	v(\bm{\theta}) = \ln \, f_X(\bm{x} \vert \bm{\theta}).
\end{align}
Based on the MLL, the auxiliary matrices defined entrywise as
\begin{align} \label{eq_A}
	[\bm{A}_{\bm{\theta}_0} ]_{ij} = \mathbb{E}_p \left\{  \frac{\partial^2 v}{\partial \theta_i \partial \theta_j}  \Bigg\vert _{\bm{\theta}=\bm{\theta}_0} \right\}
\end{align}
and
\begin{align} \label{eq_B}
	[\bm{B}_{\bm{\theta}_0} ]_{ij} = \mathbb{E}_p \left\{  \frac{\partial v}{\partial \theta_i}  \frac{\partial v}{\partial \theta_j} \Bigg\vert _{\bm{\theta}=\bm{\theta}_0} \right\}
\end{align}
are of special \blue{interest}. {In the case of correctly specified models, the matrices defined in \eqref{eq_A} and \eqref{eq_B} are equivalent to each other (up to their sign) and correspond to the \emph{Fisher Information Matrix} (FIM).}
\end{definition}

The MCRB can now be defined based on the preceding definitions by noting that, in the misspecified case, matrices \eqref{eq_A} and \eqref{eq_B} are no longer equivalent. Instead, their {non-equality} is central to the MCRB.

\begin{definition}[Misspecified Cram\'{e}r-Rao Bound] \label{def_MCRB}
Let $f_X(\bm{x} \vert \bm{\theta})$ be misspecified with respect to $p_X(\bm{x})$. Let $\hat{\bm{\theta}}$ be an MS-unbiased estimator of $\bm{\theta}_0$ following \Cref{def_MU}. Consider additionally the error covariance matrix of $\hat{\bm{\theta}}$ given by
\begin{align} \label{eq_error_covar}
	\bm{C}_p(\hat{\bm{\theta}}, \bm{\theta}_0) = \mathbb{E}_p\{ (\hat{\bm{\theta}} - \bm{\theta}_0) (\hat{\bm{\theta}} - \bm{\theta}_0)\trans \}.
\end{align} 
If the matrix $\bm{A}_{\bm{\theta}_0}$ defined entrywise as in \eqref{eq_A} is invertible, then
\begin{align} \label{eq_MCRB}
	\bm{C}_p (\hat{\bm{\theta}}, \bm{\theta}_0) {\succeq} \bm{A}_{\bm{\theta}_0}^{-1} \bm{B}_{\bm{\theta}_0} \bm{A}_{\bm{\theta}_0}^{-1} \overset{\scriptscriptstyle\Delta}{=} \text{MCRB}(\bm{\theta}_0),
\end{align}
where a matrix inequality of the form $\bm{U} {\succeq} \bm{V}$ signifies that $\bm{U} - \bm{V}$ is positive semidefinite and $\text{MCRB}(\bm{\theta}_0)$ is the \emph{Misspecified Cram\'{e}r-Rao Bound}.
\end{definition}

To compute the MCRB, a choice of true and misspecified models must be made and used in tandem with these definitions. Note that \eqref{eq_MCRB} depends explicitly on the \blue{PTP}. In general, computing the \blue{PTP} analytically is challenging. The following theorem allows this to be circumvented.

\begin{theorem}[Mismatched Maximum Likelihood Estimator] \label{th_MMLE}
The quantity 
\begin{align} \label{eq_MMLE}
	\hat{\bm{\theta}}_\text{MML} = \argmax_{\bm{\theta} \in \Theta} v(\bm{\theta})
\end{align}
is referred to as the \emph{Mismatched Maximum Likelihood Estimator} (MMLE). If regularity conditions are met, the MMLE is asymptotically (as the number of observations goes to infinity) MS-unbiased and its error covariance coincides with the MCRB.

{\emph{Proof:} {The works \cite{MCRBreal, MCRB} discuss this result in detail and refer to the original proof presented by Huber and White.}} \hfill $\blacksquare$
\end{theorem}

Based on \Cref{th_MMLE}, the MMLE is a practical alternative to the analytic computation of the {PTP} and is the method of choice in this work. We furthermore narrow our focus to the study of Gaussian models with misspecified mean, for which specialized formulations of the KLD, MLL, and MCRB can be obtained. These formulations are presented next.

\subsection{Gaussian Models with Misspecified Mean} \label{subsec_slepian}
Let $\bm{y} \in \compl^N$ be the observed data which is corrupted by zero mean, circularly symmetric, additive complex Gaussian noise $\bm{n} \in \compl^N \sim \mathcal{CN}(\bm{0}, \bm{\Sigma)}$, where {$\bm{\Sigma} \in \compl^{N \times N}$ is the noise covariance matrix}. The observed data can then be written as $\bm{y} = \bm{s} + \bm{n}$. {\iffalse{\sout{, with $\bm{s}$ and $\bm{n}$ uncorrelated.}}\fi} The vector $\bm{s} \in \compl^N$ is the true mean of the data, and so $\bm{y} \sim \mathcal{CN}(\bm{s}, \bm{\Sigma})$ defines the true PDF $p_Y(\bm{y})$. Misspecification is introduced through the assumption (be it because $\bm{s}$ is unknown or due to practical considerations) that the data $\bm{y}$ has a mean $\bm{\mu}(\bm{\theta})$ which is different from $\bm{s}$ for all values of the parameter $\bm{\theta}$. This means that the data is assumed to obey {$\bm{y} \sim \mathcal{CN}(\bm{\mu}(\bm{\theta}), \bm{\Sigma})$}, which in turn defines the misspecified PDF $f_Y(\bm{y} \vert \bm{\theta})$.

In this scenario, the KLD from $f_{Y \vert \bm{\theta}}$ to $p_Y$ takes the well-known form 
\begin{align} \label{eq_kld}
	{D(p_Y \Vert f_{Y\vert \bm{\theta}}) = (\bm{s} - \bm{\mu}(\bm{\theta}))\herm \bm{\Sigma}^{-1} (\bm{s} - \bm{\mu}(\bm{\theta}))}.
\end{align}
Similarly, the MLL is given by
\begin{equation}
\begin{split}
	v(\bm{\theta}) = & - \ln(\pi^N \text{det}\{ \bm{\Sigma} \})  \\
	& - (\bm{y} - \bm{\mu}(\bm{\theta}))\herm \bm{\Sigma}^{-1} (\bm{y} - \bm{\mu}(\bm{\theta})).
\label{eq_MLL}
\end{split}
\end{equation}
Applying the definitions from \Cref{subsec_general_defs} to this Gaussian model scenario, a generalization of the Slepian formula \cite{stoica} is obtained. {We employ the formulation shown in \cite{richmond2013parameter}, dubbed} \emph{Deterministic Generalized Slepian Formula} in \cite{gSlepian}, {noting that the resulting bound applies to any MS-unbiased estimator \cite{vuong1986cramer ,MCRB, MCRBreal, MCRBcomplex}}. The expressions are reworked for the specific case where the model parameters are real, i.e. the mean is a function $\bm{\mu} : \real^d \to \compl^N$.

\begin{theorem}[Generalized Slepian Formulae] \label{th_slepian}
	Assume the regularity conditions hold so that the MCRB is given by $\text{MCRB}(\bm{\theta}_0) \overset{\scriptscriptstyle\Delta}{=} \bm{A}_{\bm{\theta}_0}^{-1} \bm{B}_{\bm{\theta}_0} \bm{A}_{\bm{\theta}_0}^{-1}$. Consider a Gaussian model whose true mean $\bm{s} \in \compl^N$ is misspecified as $\bm{\mu}(\bm{\theta}) \in \compl^N$ with $\bm{\theta} \in \real^d$. The matrices $\bm{A}_{\bm{\theta}_0}$ and $\bm{B}_{\bm{\theta}_0}$ are defined entrywise by the generalized Slepian formulae
\begin{equation}
\begin{split}
	\frac{[\bm{A}_{\bm{\theta}_0} ]_{ij}}{2} = 
	\text{Re} \Bigg\{ &
	- \frac{\partial \bm{\mu}(\bm{\theta})}{\partial \theta_i} \herm \bm{\Sigma}^{-1}  \frac{\partial \bm{\mu}(\bm{\theta})}{\partial \theta_j}  \\
	& +\frac{\partial^2 \bm{\mu}(\bm{\theta})}{\partial \theta_i \partial \theta_j} \herm \bm{\Sigma}^{-1} ( \bm{s} - \bm{\mu}(\bm{\theta}) ) \Bigg\vert _{\bm{\theta}=\bm{\theta}_0} \Bigg\}
\end{split}
\label{eq_hess}
\end{equation}
and
\begin{equation}
\begin{split}
	{\frac{[\bm{B}_{\bm{\theta}_0} ]_{ij}}{2} = \text{Re} \Bigg\{} &
	{\frac{\partial \bm{\mu}(\bm{\theta})}{\partial \theta_i} \trans \bm{\Sigma}^{-1} \bm{\Delta}(\bm{\theta})^* \bm{\Delta}(\bm{\theta})\herm \bm{\Sigma}^{-1} \frac{\partial \bm{\mu}(\bm{\theta})}{\partial \theta_j}} \\
	& {+\frac{\partial \bm{\mu}(\bm{\theta})}{\partial \theta_i} \herm \bm{\Sigma}^{-1} \bm{\Delta}(\bm{\theta}) \bm{\Delta}(\bm{\theta})\herm \bm{\Sigma}^{-1}  \frac{\partial \bm{\mu}(\bm{\theta})}{\partial \theta_j}}\\
	& {+\frac{\partial \bm{\mu}(\bm{\theta})}{\partial \theta_i} \herm \bm{\Sigma}^{-1} \frac{\partial \bm{\mu}(\bm{\theta})}{\partial \theta_j} \Bigg\vert _{\bm{\theta}=\bm{\theta}_0}	\Bigg\},}
\end{split}
\end{equation}
{where $\bm{\Sigma} \in \compl^{N \times N}$ is the noise covariance matrix} shared by the true and misspecified models {and $\bm{\Delta}(\bm{\theta}) = \bm{s} - \bm{\mu}(\bm{\theta})$}. 

\emph{Proof:} {Different derivations are presented in \cite{richmond2013parameter}, \cite{mennad2018slepian}, and \cite{gSlepian}}. \hfill $\blacksquare$
\end{theorem}

Note that, if $\bm{\mu}(\bm{\tilde{\theta}}) = \bm{s}$ for some value $\bm{\tilde{\theta}}$ of $\bm{\theta}$, the model is no longer misspecified and all terms involving the difference {$\bm{\Delta}(\bm{\theta}) = \bm{s} - \bm{\mu}(\bm{\theta})$} become $\bm{0}$. In this case, $\bm{B}_{\bm{\theta}_0} = -\bm{A}_{\bm{\theta}_0}$, both Slepian formulae correspond to the correctly specified FIM, and the MCRB is equal to the classical CRB.

As the goal is to apply these results to true and misspecified models related to scalar wave phenomena, we discuss the basics of scalar wave fields and wave propagation models next. In the remainder of this work, the upcoming propagation models will be referred to simply as \emph{models}, as they describe how the means $\bm{\mu}$ and $\bm{s}$ of the misspeficied and true statistical models, respectively, are constructed.

%% file: sections/models.tex
\section{Scalar Wave Field Models} \label{section_models}
Wave phenomena can be modeled through the Helmholtz equation \cite{fwi_original, simulation, Pade}, which in the presence of sources takes the inhomogeneous form
\begin{align}
	{\nabla^2 \psi_\text{t}(\omega, \bm{x}) + k^2(\omega, \bm{x}) \psi_\text{t}(\omega, \bm{x}) = -s(\omega, \bm{x}).}
	\label{eq_hh}
\end{align}
In \eqref{eq_hh}, {$\bm{x} \in \real^2$} denotes a position vector, $\omega$ is the angular frequency, {$\psi_\text{t} : \real \times \real^2 \to \compl$} is the  \emph{total field} in a \emph{region of interest} (ROI), {$k: \real \times \real^2 \to \compl$} is the wavenumber, and {$s: \real \times \real^2 \to \compl$} is a harmonic excitation source. The wavenumber can be decomposed into 
\begin{align}
	k(\omega, \bm{x}) = \frac{\omega}{c(\bm{x})} - \jmath \beta({\omega, \bm{x}})
\end{align}
where {$c: \real^2 \to \real$} assigns a propagation speed to each point in space and {$\beta: \real \times \real^2 \to \real$} is a frequency and space dependent attenuation factor. 

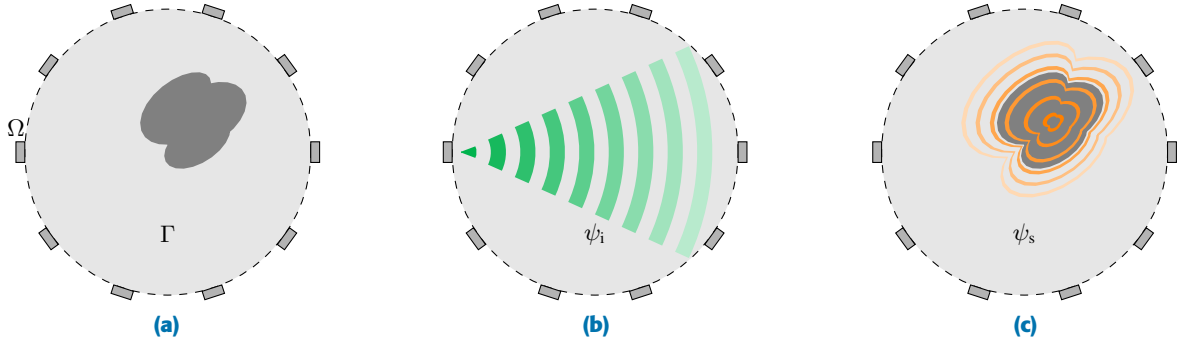
\begin{figure*}
	\centering
	\begin{subfigure}{0.32\textwidth}
		\hspace{0.52cm}
		\input{TikZ_sketches/setup}
		\caption{}
		\label{fig_scenario}
	\end{subfigure}%
	\begin{subfigure}{0.32\textwidth}
		\hspace{0.72cm}
		\input{TikZ_sketches/incident}
		\caption{}
		\label{fig_incident}
	\end{subfigure}%
	\begin{subfigure}{0.32\textwidth}
		\hspace{0.72cm}
		\input{TikZ_sketches/scattered}
		\caption{}
		\label{fig_scattered}
	\end{subfigure}
	\caption{Illustration of the usage of contrast sources. In \Cref{fig_scenario}, an illustration of a generic tomography setup is shown. A transducer ring is placed inside a medium that extends infinitely in all directions. Within the medium, a darker gray color represents a wavenumber that differs from that of the background. The total field within the medium can be expressed in terms of two distinct components. The incident field is shown in \Cref{fig_incident}, where it is clear that the inhomogeneity is no longer present and the excitation produced by a transducer propagates in the homogeneous medium. In \Cref{fig_scattered}, the inhomogeneities are shown behaving as sources producing a field that also propagates through the homogeneous background medium characterized by $\boldsymbol{k_0}$.}
	\label{fig_fields}
\end{figure*}

It is important to note that the properties of $\beta$ depend on the application: when dealing with ultrasound and human tissue, $\beta$ is approximately linearly dependent on the angular frequency $\omega$ \cite{atten_tissue}; however, a nonlinear dependence on $\omega$ may be present when dealing with ultrasound in metals \cite{metals1, metals2}. {Assuming a linear dependence on $\omega$ for simplicity,} the wavenumber $k$ can be rewritten as
\begin{align}
	k(\bm{x}) = \omega \left( \frac{1}{c(\bm{x})} - \jmath \beta({\bm{x}}) \right),
\end{align}
with an attenuation factor $\beta$ that depends only on the position $\bm{x}$ after factoring $\omega$.

{Problems involving \eqref{eq_hh} are categorized into forward problems where $k(\bm{x})$ and $s(\omega, \bm{x})$ are known and $\psi_\text{t}(\omega, \bm{x})$ is sought after, and inverse problems in which $k(\bm{x})$ is estimated given $s(\omega, \bm{x})$ and samples of $\psi_\text{t}(\omega, \bm{x})$.} A common solution approach {for inverse problems} is to introduce a contrast term so that only deviations with respect to a reference value of $k$ are considered \cite{kaczmarz, fwi_original, fwi_new, simulation, CSUCT, Pade}. One possible formulation of the contrast is
\begin{align}
	\hat{\gamma}(\omega, \bm{x}) = k^2(\omega, \bm{x}) - k_0^2,
\end{align}
where {$\hat{\gamma}: \real \times \real^2 \to \compl$} is referred to as the wavenumber contrast, and $k_0 \in \compl$ is a complex scalar denoting the background wavenumber from which $k$ deviates. By recalling the assumption that $k$ depends linearly on $\omega$, the quantity
\begin{align}
	\gamma(\bm{x}) = \left[ \left( \frac{1}{c(\bm{x})} - \jmath \beta({\bm{x}}) \right)^2 - \left( \frac{1}{c_0} - \jmath \beta_0 \right)^2 \right]
	\label{eq_flatcontrast}
\end{align}
can be defined, allowing the contrast $\hat{\gamma}$ to be reformulated as 
\begin{align}
	\hat{\gamma}(\omega, \bm{x}) = \omega^2 \gamma(\bm{x}).
\end{align}
Due to this relationship, the quantity {$\gamma: \real^2 \to \compl$} as given by \eqref{eq_flatcontrast} is referred to as the \emph{frequency flat contrast}. The frequency flat contrast allows \eqref{eq_hh} to be rewritten in the form
\begin{align}
	{\left\{ \nabla^2 + k_0^2 \right\} \psi_\text{t}(\omega, \bm{x}) = -s(\omega, \bm{x}) - \omega^2 \gamma(\bm{x}) \psi_\text{t}(\omega, \bm{x}),}
	\label{eq_hh_contrast}
\end{align}
where $\nabla^2$ and $k_0^2$ have been grouped together to highlight their behavior as a differential operator acting on $\psi_\text{t}$. Crucially, this operator now depends on $k_0$ rather than the spatially varying $k$. Paired with the Sommerfeld radiation condition, this allows \eqref{eq_hh_contrast} to be solved by using the Green's function associated to the differential equation. The Green's function {$G: \real \times \real^2 \times \real^2 \to \compl$} is given by
\begin{align}
	G(\omega, \bm{x}, \bm{x}') = -\frac{\jmath}{4} H_0^{(2)}(k_0 \Vert \bm{x} - \bm{x}' \Vert_2),
	\label{eq_hankel}
\end{align}
where $H_0^{(2)}$ is the zeroth order Hankel function of the second kind, and $\bm{x}, \bm{x}' \in \real^2$ denote locations in the ROI. 

Using the Green's function in \eqref{eq_hankel}, the total field can be expressed as
\begin{align}
	{\psi_\text{t}(\bm{x}) = \psi_\text{i}(\bm{x}) + \psi_\text{s}(\bm{x}),}
	\label{eq_total}
\end{align}
with subindices $\rm{i}$ and $\rm{s}$ referring to \emph{incident} and \emph{scattered} fields, respectively, and where dependence on $\omega$ has been omitted. In turn, these fields are given by
\begin{align}
	{\psi_\text{i}(\bm{x}) = \int_\Omega G(\bm{x}, \bm{x}') s(\bm{x}') \, d\bm{x}'}
	\label{eq_incident}
\end{align}
and
\begin{align}
	{\psi_\text{s}(\bm{x}) = \omega^2 \int_\Gamma G(\bm{x}, \bm{x}') \gamma(\bm{x}') \psi_\text{t}(\bm{x}') \, d\bm{x}',}
	\label{eq_scattered}
\end{align}
where $\Omega$ refers to the {support of an excitation source or transmitter} and $\Gamma$ is the ROI. As \eqref{eq_incident} does not depend on $\gamma$, the incident field $\psi_\text{i}$ is considered to be known or computed ahead of time. Equations \eqref{eq_incident} and \eqref{eq_scattered} can be interpreted as decomposing the total field into the contributions from two kinds of sources: the {original sources} that excite a medium where $\gamma$ is everywhere zero, and regions in the medium where $\gamma \neq 0$ which behave as additional sources. This is illustrated in \Cref{fig_fields}. {Note that, since the contrast $\gamma$ behaves as a source in \eqref{eq_scattered}, the scattered field scales with the magnitude of the contrast. 

{Equations \eqref{eq_total}-\eqref{eq_scattered} can be discretized through different choices of sampling grids and integral approximations. These choices result in distinct forward models, as discussed next.}

\subsection{{Helmholtz Model}}

{Consider an ROI sampled with resolution $\Delta_x$ along the horizontal axis and $\Delta_z$ along the vertical axis so as to yield $N_z \cdot N_x$ points on a regular 2D grid.} {After discretization, we represent the total and incident fields in the ROI at frequency $\omega$ and the frequency flat contrast as vectors of the form $\bm{\psi}_\text{t}, \bm{\psi}_\text{i}, \bm{\gamma} \in \compl^{N_z \cdot N_x}$.} After choosing a numerical integration scheme {for \eqref{eq_scattered}}, a matrix $\Groi \in \compl^{N_z \cdot N_x \times N_z \cdot N_x}$ is defined so that the total field in the ROI defined in \eqref{eq_total} can be approximated as
\begin{align}
	{\ftot = \finc + \Groi (\bm{\gamma} \circ \ftot) = \finc + \Groi \diag{\bm{\gamma}} \ftot,}
	\label{eq_ftot}
\end{align}
where the factor $\omega^2$ has been absorbed into $\Groi$, $\circ$ denotes the Hadamard or elementwise product, {and $\diag{\cdot}$ yields a diagonal matrix whose entries correspond to the given argument}. Furthermore, in the absence of self-interference and considering point-like receiver and transmitter elements, the field observed at a receiver is equivalent to the scattered field given by \eqref{eq_scattered} when evaluated at the coordinates of said receiving element. For an array with $M_R$ receiving elements, a matrix $\Garray \in \compl^{M_R \times N_z \cdot N_x}$ can be defined so that \eqref{eq_scattered} is approximated by
\begin{align}
	\fsc = \Garray \diag{\ftot} \bm{\gamma}
	\label{eq_gamma}
\end{align}
by evoking the commutativity of the Hadamard product to swap $\ftot$ and $\bm{\gamma}$, and where $\fsc \in \compl^{M_R}$ is the scattered field measured at the receiver array at frequency $\omega$. Data comprising $N_{\omega}$ frequency bins of interest and $M_T$ transmitting elements can be obtained by using \eqref{eq_ftot} and \eqref{eq_gamma} $N_\omega \cdot M_T$ times. {We refer to} \eqref{eq_ftot} {and} \eqref{eq_gamma} {as the \emph{Helmholtz model} throughout this work.}

Note that, in the special case where $\Groi$ and $\Garray$ are constructed by simply taking samples of the Green's function, \eqref{eq_ftot} and \eqref{eq_gamma} correspond to the \emph{Foldy-Lax} model \cite{foldy1945multiple, lax1951multiple, crb_mono1, crb_mono2, crb_resolved}. This is equivalent to considering the contrast $\gamma$ as being caused by point scatterers represented by a sum of delta functions in space. Substituting such a contrast term into \eqref{eq_total} and \eqref{eq_scattered} directly yields \eqref{eq_ftot} and \eqref{eq_gamma}.
 
\subsection{{Born Model}}
{In both forward and inverse problems involving the Helmholtz equation \eqref{eq_hh}}, the largest computational effort is spent in solving \eqref{eq_ftot} for $\ftot$. A common solution is to make the simplifying assumption that, at the scatterers, the total field on the {right-hand side} {of \eqref{eq_ftot}} is equivalent to the incident field. {The justification behind this is that, if the contrast $\bm{\gamma}$ is small, then the incident field dominates the scattered field and $\ftot = \finc + \fsc \approx \finc$}. {This is known as the first order Born approximation, here referred to as the \emph{Born model}, and it can be interpreted as ignoring the interactions among the scatterers and therefore omitting multiple scattering.} A full data set can then be obtained by computing
\begin{align}
	{\fsc \approx \bm{\psi}_\text{B} = \Garray \diag{\finc} \bm{\gamma}}
	\label{eq_born}
\end{align}
for all combinations of frequency and transmitter indices.

\subsection{Delay Model}
A further simplification can be made by considering point-like scatterers and modifying the Green's function in two ways. First, the Green's function can be approximated as
\begin{align}
	G(\bm{d}) \approx -\frac{\jmath}{4} \sqrt{ \frac{2}{\pi k_0 \Vert \bm{d} \Vert_2 }} \exp \left( -\jmath \left( k_0 \Vert \bm{d} \Vert_2 - \frac{\pi}{4} \right) \right)
	\label{eq_hankel_asymp}
\end{align}
for large distance values $\Vert \bm{d} \Vert_2 = \Vert \bm{x} - \bm{x}' \Vert_2$. Next, the $\sqrt{k_0 \Vert \bm{d} \Vert_2}$ in the denominator is dropped. This amounts to ignoring the \emph{beam spread} as waves travel away from their source. {If we assume the background wavenumber $k_0$ to be real, we assume the medium to be non-attenuating.} This turns \eqref{eq_hankel_asymp} into the transfer function of a time delay. The overall forward model subsequently turns into a sum of copies of a transmitted waveform, each of which has been delayed depending on the geometry concerning the sensors and the scatterers, and each copy is additionally scaled based on the value of $\gamma$.

These simplifications can be used to rewrite the forward model as follows. {As the transmitters are considered to be point-like, the excitation $s(\omega, \bm{x})$ is of the form $a(\omega) \delta(\bm{x} - \bm{x}_t)$ for the $t^\text{th}$ transducer located at $\bm{x}_t$.} The amplitude {$a: \real \to \compl$} corresponds to the spectral amplitude or Fourier coefficient at angular frequency $\omega$ of an excitation pulse shape $p(t)$ with spectrum $P(\omega)$. Scaled, delayed copies of this pulse shape are measured at the receivers, one for each scatterer. The scattered field observed at a fixed receiver after a medium with $U$ scatterers is excited by a transmitter is then approximately given by
\begin{align}
	{\psi_{\text{s}}(\omega) \approx \psi_\text{D} = P(\omega) \sum_{u=1}^U q_u \exp ( -\jmath \omega \tau(\bm{x}_u)),}
	\label{eq_simple}
\end{align}
where $q_u \in \compl$ is the \emph{scattering coefficient} of the $u^\text{th}$ scatterer and $\tau(\bm{x}_u)$ is its associated time delay, noting that this time delay depends both on the scatterer's location $\bm{x}_u$ and the location of the transmitter and receiver. The time delay can be written as
\begin{align}
	\tau_{r, t}(\bm{x}_u) = \frac{1}{c_0} \left( \Vert \bm{x}_t - \bm{x}_u \Vert_2 + \Vert \bm{x}_u - \bm{x}_r \Vert_2 \right)
\end{align}
for a transmitter located at $\bm{x}_t$ and a receiver at $\bm{x}_r$. Such a model is employed in migration and delay and sum techniques for imaging and localization, e.g. the \emph{Synthetic Aperture Focusing Technique} \cite{SAFT} and the \emph{Total Focusing Method} \cite{TFM} in ultrasound.

\begin{figure}
	\centering
	\includegraphics[width=1\columnwidth, trim = {0cm 0cm 0cm 0cm}, clip]{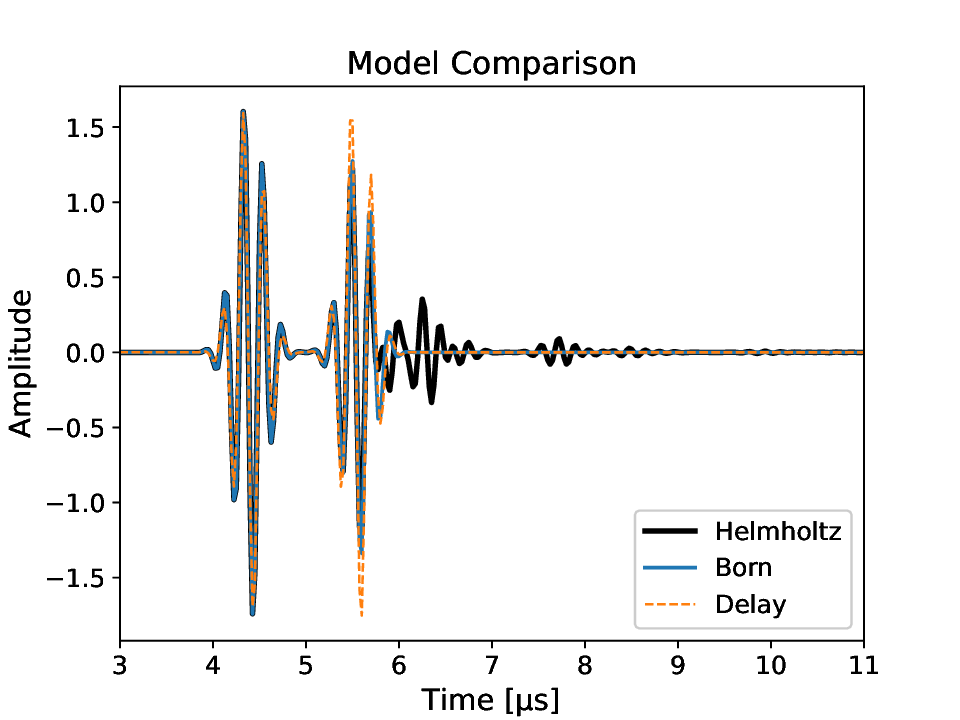}
	\caption{Scattered fields simulated via the three different models in the presence of two scatterers. Good agreement among the three is seen near the dominant echoes; however, neither the Born approximation nor the delay model exhibit multiple scattering.}
	\label{fig_model_comparison}
	\vspace{-0.5cm}
\end{figure}

A comparison among the three aforementioned models is shown in \Cref{fig_model_comparison} {in the time domain}. In this figure, the \emph{Helmholtz} model refers to fields produced via \eqref{eq_ftot} and \eqref{eq_gamma}, the \emph{Born} model corresponds to \eqref{eq_born}, and the \emph{Delay} model refers to \eqref{eq_simple}. The plot shows the scattered field observed by a point receiver when a medium containing two point scatterers is excited by a point transmitter emitting a Gabor function. The exact details of the simulation are presented in \Cref{section_simulations}. Note that the Helmholtz model, shown in black, exhibits multiple scattering which manifests as reverberation. The Born approximation, in blue, agrees with the two largest pulses that can be observed in the Helmholtz model; however, reverberation is absent. Finally, the delay model lacks reverberation and, although the amplitudes of the largest echoes are similar to those of the previous two models, there is also a discrepancy due to the simplified Green's function.

%% file: TikZ_sketches/setup.tex
\begin{tikzpicture}
\tikzstyle{every node}=[font=\small]

\pgfmathsetmacro\scale{0.39} 
\pgfmathsetmacro\pi{3.14159}

\pgfmathsetmacro\ntransducers{10} 
\pgfmathsetmacro\radius{5*\scale} 

\pgfmathsetmacro\tw{0.7*\scale} 
\pgfmathsetmacro\th{0.3*\scale} 

\draw[dashed,fill=gray!20!white] (0,0) circle (\radius - \th/2);


\node at (0, -\radius/1.8) {$\Gamma$};
\node at (-\radius - \th/2, \tw*1.2) {$\Omega$};

\foreach \t in {1,...,\ntransducers}{
	\pgfmathsetmacro\angle{360/(\ntransducers)*((\t)-1)}
	\pgfmathsetmacro\tx{\radius*cos(\angle)}
	\pgfmathsetmacro\ty{\radius*sin(\angle)}
	\draw[fill=gray!60!white, rotate around={\angle + 90:(\tx,\ty)}] ({\tx-\tw/2},{\ty-\th/2}) -- ({\tx-\tw/2},{\ty+\th/2}) 
	-- ({\tx+\tw/2},{\ty+\th/2}) -- ({\tx+\tw/2},{\ty-\th/2}) -- cycle;
}

put in a defect with a rather arbitrary contour
\pgfmathsetmacro\r{2*\scale} 
\pgfmathsetmacro\o{1*\scale} 
\fill[gray, domain=0:360, samples = 100, variable=\t] 
	({\r/2.7*(cos(0) - 0.5*cos(2*0 + 45) + 1.2*cos(5*0 - 240)) + \o},{\r/2.3*(sin(0) - 0.5*sin(2*0 + 45) + 0.8*sin(5*0- 210))+ \o}) -- 
	plot({\r/2.7*(cos(\t) - 0.5*cos(2*\t + 45) + 1.2*cos(5*\t - 240)) + \o},{\r/2.3*(sin(\t) - 0.5*sin(2*\t + 45) + 0.8*sin(5*\t - 210)) + \o})
	-- ({\r/2.7*(cos(0) - 0.5*cos(2*0 + 45) + 1.2*cos(5*0 - 240)) + \o},{\r/2.3*(sin(0) - 0.5*sin(2*0 + 45) + 0.8*sin(5*0- 210)) + \o}) 
	-- cycle;

\end{tikzpicture}

%% file: TikZ_sketches/incident.tex
\begin{tikzpicture}
\tikzstyle{every node}=[font=\small]

\pgfmathsetmacro\scale{0.39} 
\pgfmathsetmacro\pi{3.14159}

\pgfmathsetmacro\ntransducers{10} 
\pgfmathsetmacro\radius{5*\scale} 

\pgfmathsetmacro\tw{0.7*\scale} 
\pgfmathsetmacro\th{0.3*\scale} 

\draw[dashed,fill=gray!20!white] (0,0) circle (\radius - \th/2);


\node at (0, -\radius/1.8) {$\psi_\text{i}$};

\foreach \t in {1,...,\ntransducers}{
	\pgfmathsetmacro\angle{360/(\ntransducers)*((\t)-1)}
	\pgfmathsetmacro\tx{\radius*cos(\angle)}
	\pgfmathsetmacro\ty{\radius*sin(\angle)}
	\draw[fill=gray!60!white, rotate around={\angle + 90:(\tx,\ty)}] ({\tx-\tw/2},{\ty-\th/2}) -- ({\tx-\tw/2},{\ty+\th/2}) 
	-- ({\tx+\tw/2},{\ty+\th/2}) -- ({\tx+\tw/2},{\ty-\th/2}) -- cycle;
}

\pgfmathsetmacro\ba{-25} 
\pgfmathsetmacro\ta{-\ba} 
\foreach \iter in {1,...,9}{
	\pgfmathsetmacro\r{\radius*2*(\iter - 0.75)/10} 
	\pgfmathsetmacro\cx{(-\radius + \r)*cos(\ba)} 
	\pgfmathsetmacro\cy{(\r)*sin(\ba)} 
	\pgfmathsetmacro\wh{100 - (\iter-1)*9} 
	\draw[line width=2mm, green!70!blue!\wh!white] (\cx,\cy) arc (\ba:\ta:\r);
}

\end{tikzpicture}

%% file: TikZ_sketches/scattered.tex
\begin{tikzpicture}
\tikzstyle{every node}=[font=\small]

\pgfmathsetmacro\scale{0.39} 
\pgfmathsetmacro\pi{3.14159}

\pgfmathsetmacro\ntransducers{10} 
\pgfmathsetmacro\radius{5*\scale} 

\pgfmathsetmacro\tw{0.7*\scale} 
\pgfmathsetmacro\th{0.3*\scale} 

\draw[dashed,fill=gray!20!white] (0,0) circle (\radius - \th/2);


\node at (0, -\radius/1.8) {$\psi_\text{s}$};

\foreach \t in {1,...,\ntransducers}{
	\pgfmathsetmacro\angle{360/(\ntransducers)*((\t)-1)}
	\pgfmathsetmacro\tx{\radius*cos(\angle)}
	\pgfmathsetmacro\ty{\radius*sin(\angle)}
	\draw[fill=gray!60!white, rotate around={\angle + 90:(\tx,\ty)}] ({\tx-\tw/2},{\ty-\th/2}) -- ({\tx-\tw/2},{\ty+\th/2}) 
	-- ({\tx+\tw/2},{\ty+\th/2}) -- ({\tx+\tw/2},{\ty-\th/2}) -- cycle;
}

\pgfmathsetmacro\o{1*\scale} 
\foreach \iter in {1,...,3}{
	\pgfmathsetmacro\r{2*\scale + (3-\iter + 0.5)/2*\scale} 
	\pgfmathsetmacro\wh{30 + (\iter-1)*20 }
	\fill[orange!\wh!white, domain=0:360, samples = 100, variable=\t] 
		({\r/2.7*(cos(0) - 0.5*cos(2*0 + 45) + 1.2*cos(5*0 - 240)) + \o},{\r/2.3*(sin(0) - 0.5*sin(2*0 + 45) + 0.8*sin(5*0- 210))+ \o}) -- 
		plot({\r/2.7*(cos(\t) - 0.5*cos(2*\t + 45) + 1.2*cos(5*\t - 240)) + \o},{\r/2.3*(sin(\t) - 0.5*sin(2*\t + 45) + 0.8*sin(5*\t - 210)) + \o})
		-- ({\r/2.7*(cos(0) - 0.5*cos(2*0 + 45) + 1.2*cos(5*0 - 240)) + \o},{\r/2.3*(sin(0) - 0.5*sin(2*0 + 45) + 0.8*sin(5*0- 210)) + \o}) 
		-- cycle;

	\pgfmathsetmacro\r{2*\scale + (3-\iter + 0.2)/2*\scale} 
	\fill[gray!20!white, domain=0:360, samples = 100, variable=\t] 
		({\r/2.7*(cos(0) - 0.5*cos(2*0 + 45) + 1.2*cos(5*0 - 240)) + \o},{\r/2.3*(sin(0) - 0.5*sin(2*0 + 45) + 0.8*sin(5*0- 210))+ \o}) -- 
		plot({\r/2.7*(cos(\t) - 0.5*cos(2*\t + 45) + 1.2*cos(5*\t - 240)) + \o},{\r/2.3*(sin(\t) - 0.5*sin(2*\t + 45) + 0.8*sin(5*\t - 210)) + \o})
		-- ({\r/2.7*(cos(0) - 0.5*cos(2*0 + 45) + 1.2*cos(5*0 - 240)) + \o},{\r/2.3*(sin(0) - 0.5*sin(2*0 + 45) + 0.8*sin(5*0- 210)) + \o}) 
		-- cycle;
}

\pgfmathsetmacro\r{2*\scale} 
\fill[gray, domain=0:360, samples = 100, variable=\t] 
	({\r/2.7*(cos(0) - 0.5*cos(2*0 + 45) + 1.2*cos(5*0 - 240)) + \o},{\r/2.3*(sin(0) - 0.5*sin(2*0 + 45) + 0.8*sin(5*0- 210))+ \o}) -- 
	plot({\r/2.7*(cos(\t) - 0.5*cos(2*\t + 45) + 1.2*cos(5*\t - 240)) + \o},{\r/2.3*(sin(\t) - 0.5*sin(2*\t + 45) + 0.8*sin(5*\t - 210)) + \o})
	-- ({\r/2.7*(cos(0) - 0.5*cos(2*0 + 45) + 1.2*cos(5*0 - 240)) + \o},{\r/2.3*(sin(0) - 0.5*sin(2*0 + 45) + 0.8*sin(5*0- 210)) + \o}) 
	-- cycle;

\foreach \iter in {1,...,3}{
	\pgfmathsetmacro\r{2*\scale - (\iter*2/3 - 0.4)*\scale} 
	\pgfmathsetmacro\wh{70 + (\iter)*10 }
	\fill[orange!\wh!white, domain=0:360, samples = 100, variable=\t] 
		({\r/2.7*(cos(0) - 0.5*cos(2*0 + 45) + 1.2*cos(5*0 - 240)) + \o},{\r/2.3*(sin(0) - 0.5*sin(2*0 + 45) + 0.8*sin(5*0- 210))+ \o}) -- 
		plot({\r/2.7*(cos(\t) - 0.5*cos(2*\t + 45) + 1.2*cos(5*\t - 240)) + \o},{\r/2.3*(sin(\t) - 0.5*sin(2*\t + 45) + 0.8*sin(5*\t - 210)) + \o})
		-- ({\r/2.7*(cos(0) - 0.5*cos(2*0 + 45) + 1.2*cos(5*0 - 240)) + \o},{\r/2.3*(sin(0) - 0.5*sin(2*0 + 45) + 0.8*sin(5*0- 210)) + \o}) 
		-- cycle;

	\pgfmathsetmacro\r{2*\scale - (\iter*2/3 - 0.25)*\scale} 
	\fill[gray, domain=0:360, samples = 100, variable=\t] 
		({\r/2.7*(cos(0) - 0.5*cos(2*0 + 45) + 1.2*cos(5*0 - 240)) + \o},{\r/2.3*(sin(0) - 0.5*sin(2*0 + 45) + 0.8*sin(5*0- 210))+ \o}) -- 
		plot({\r/2.7*(cos(\t) - 0.5*cos(2*\t + 45) + 1.2*cos(5*\t - 240)) + \o},{\r/2.3*(sin(\t) - 0.5*sin(2*\t + 45) + 0.8*sin(5*\t - 210)) + \o})
		-- ({\r/2.7*(cos(0) - 0.5*cos(2*0 + 45) + 1.2*cos(5*0 - 240)) + \o},{\r/2.3*(sin(0) - 0.5*sin(2*0 + 45) + 0.8*sin(5*0- 210)) + \o}) 
		-- cycle;
}

\end{tikzpicture}

%% file: sections/MCRB.tex
\section{Misspecification of Multiple Scattering}
{We now apply the MCRB from \Cref{section_mcrbdef} to study the achievable estimation performance in localization tasks involving the scalar wave phenomena from \Cref{section_models}. Misspecification occurs when estimating parameters based on a simplified model such as the Born or delay model, ignoring the presence of multiple scattering in the data.} 

We focus on the scenario in which the model follows a circularly symmetric complex Gaussian distribution whose mean is misspecified. We highlight the interpretability of the generalized Slepian formulae presented in \Cref{subsec_slepian}, which explicitly account for the mismatch between the true and assumed models. More concretely, when applying the generalized Slepian formulae in the present application, the true and misspecified means correspond to different scalar wave propagation models, and the MCRB is explicitly formulated in terms of their difference.

\subsection{Misspecification of the Propagation Model} \label{subsec_missp}
{We contribute to the discussion on the impact of multiple scattering on estimation performance by analyzing the model mismatch via the MCRB.} An MCRB that evaluates the impact of the misspecification of multiple scattering can be obtained by combining the models in \Cref{section_models} and the generalized Slepian formulae in \Cref{th_slepian}. In order to account for multiple scattering, the true mean is taken to be {$\bm{s} = \bm{\psi}_\text{s} \in \compl^{N_\omega \cdot M_R \cdot M_T}$} as given by the solution to the Helmholtz equation obtained by solving equations \eqref{eq_ftot} and \eqref{eq_gamma} for all $N_\omega$ frequencies and $M_T$ transmitters. 

{This true model depends on $\bm{\gamma}$, which has the locations $\bm{x}_u$ of $U$ point-like scatterers as its support and assigns the scatterers a wavenumber contrast $\gamma_u = \gamma(\bm{x}_u)$ with respect to the background medium. For the sake of clarity, let us define a \emph{true parameter} $\bm{\phi}$ containing the aforementioned scatterer locations $\bm{x}_u$ and their corresponding wavenumber contrasts $\gamma(\bm{x}_u)$. We can then write the true mean as $\bm{s}(\bm{\phi})$. The model is additionally corrupted by circularly symmetric complex Gaussian noise $\bm{n} \sim \mathcal{CN}(\bm{0}, \bm{\Sigma})$.}

{The misspecified mean is chosen as the delay-based model from \eqref{eq_simple} after discretization in the frequency domain, and the same noise $\bm{n}$ as before is considered. This results in a model} {$\bm{\mu}(\bm{\theta}) = \bm{\psi}_\text{D} \in \compl^{N_\omega \cdot M_R \cdot M_T}$}. {The parameter $\bm{\theta}$ of the misspecified model describes the locations $\bm{x}_u$ of the scatterers just as $\bm{\phi}$ does, but instead of wavenumber contrasts, a complex scattering coefficient $q_u = a_u \exp(\jmath \theta_u)$ characterized by a real-valued amplitude $a_u$ and phase $\theta_u$ is assigned to each scatterer as in \eqref{eq_simple}.} 

{Crucially, the parameter spaces of the true and misspecified models are overall distinct and intersect only in the subspace of the location parameters. In the context of model misspecification, even if the parameter $\bm{\phi}$ of the true model $\bm{s}$ is known, the assumed model $\bm{\mu}$ cannot in general be evaluated at $\bm{\phi}$, e.g. when $\bm{\phi}$ and $\bm{\theta}$ are of different dimensions. Even in the case when $\bm{\mu}(\bm{\phi})$ is properly defined, it has no connection to $\bm{s}(\bm{\phi})$ as the physical meaning of the parameter spaces is different. The scattering coefficients $q_u$ in $\bm{\theta}$ represent the combined effect of the corresponding wavenumber contrast $\gamma(\bm{x}_u)$, the Green's function, and beam spread, and attribute them to a single scalar acting on the incident field. This is a large simplification of the true physical phenomenon. As an added consequence, the locations $\bm{x}_u$ of the scatterers according to the assumed model $\bm{\mu}$ generally differ from those of the true model.}

{In addition to the previous observation, recall that the MCRB provides a lower bound for the error covariance matrix $\bm{C}_p(\hat{\bm{\theta}}, \bm{\theta}_0)$. That is to say, the MCRB describes the variance of MS-unbiased estimators around the PTP $\bm{\theta}_0$. The MCRB does not, however, compare the estimated parameter $\hat{\bm{\theta}}$ against the true parameter $\bm{\phi}$. Such a comparison is difficult, since the quantities $\bm{\phi} - \bm{\theta}_0$ and $\bm{\phi} - \hat{\bm{\theta}}$, as well as the \emph{Mean Squared Error} (MSE), can only be well-defined on intersecting subspaces of the parameter spaces. Furthermore, there is no general framework with which the difference between the true and pseudo-true parameters can be studied \cite{MCRBreal, MCRB}. In the special cases where the parameter difference vector $\bm{r} = \bm{\phi} - \bm{\theta}_0$ can be defined, as is the case in our application, a connection can be drawn between the MCRB and the MSE as discussed next.}

{Throughout the remainder of this work, the scenario with $U = 2$ point scatterers is considered. In this case, $\bm{\theta} = [\bm{x}_1\trans, \bm{x}_2\trans, a_1, a_2, \theta_1, \theta_2]\trans$ and $\bm{\phi} = [\bm{x}_1\trans, \bm{x}_2\trans, \vert \gamma_1 \vert, \vert \gamma_2 \vert, \arg(\gamma_1), \arg(\gamma_2)]\trans$, with $\bm{x}_u = [x_u, z_u]\trans$ and $\gamma_u = \vert \gamma_u \vert \arg(\gamma_u) = \gamma(\bm{x}_u)$. Letting $\bm{\theta}^s = [\bm{x}_1\trans, \bm{x}_2\trans]\trans$ and $\bm{\phi}^s = [\bm{x}_1\trans, \bm{x}_2\trans]\trans$, with $(\cdot)^s$ as the projection operator onto the subspace of location parameters, the difference vector $\bm{r} = \bm{\phi}^s - \bm{\theta}_0^s$ between the true and pseudo-true parameters can be properly defined. The MSE of an MS-unbiased estimator $\hat{\bm{\theta}}$ with PTP $\bm{\theta}_0$ and true parameter $\bm{\phi}$ on the subspace denoted by $(\cdot)^s$ can then be written as \cite{MCRB}}
\begin{equation}
\begin{split} \label{eq_MSE}
	\text{MSE}(\hat{\bm{\theta}}^s, \bm{\phi}^s) & = \mathbb{E}_p \{ (\hat{\bm{\theta}}^s - \bm{\phi}^s)(\hat{\bm{\theta}}^s - \bm{\phi}^s)\trans \}\\
	& = \mathbb{E}_p \{ (\hat{\bm{\theta}}^s - \bm{\theta}_0^s)(\hat{\bm{\theta}}^s - \bm{\theta}_0^s)\trans\} + \bm{rr}\trans.
\end{split}
\end{equation}
{Due to \eqref{eq_MCRB} and \eqref{eq_MSE}, it follows that}
\begin{align} \label{eq_difference}
	{\text{MSE}(\hat{\bm{\theta}}^s, \bm{\phi}^s) - \text{MCRB}(\bm{\theta}_0^s) \succeq \bm{rr}\trans.}
\end{align}

{We highlight that, in the context of misspecification, \eqref{eq_MSE} is not to be understood as a bias-covariance decomposition. Instead, since the error covariance is defined with respect to the PTP as in \eqref{eq_error_covar}, a proper bias-covariance decomposition pertains to the \emph{Misspecified Mean squared Error} (MMSE) and is given by}
\begin{equation}
\begin{split} \label{eq_bias}
	\text{MMSE}(\hat{\bm{\theta}}, \bm{\theta}_0) & = \mathbb{E}_p\{ (\hat{\bm{\theta}} - \bm{\theta}_0)(\hat{\bm{\theta}} - \bm{\theta}_0) \}\\
	& = \mathbb{E}_p \{ (\hat{\bm{\theta}} - \bar{\bm{\theta}})(\hat{\bm{\theta}} - \bar{\bm{\theta}})\trans \} + \bm{bb}^T,
\end{split}
\end{equation}
{where $\bar{\bm{\theta}} = \mathbb{E}_p\{ \hat{\bm{\theta}} \}$ and with $\bm{b} = \bm{\theta}_0 - \bar{\bm{\theta}}$ as the bias. When $\bar{\bm{\theta}} = \bm{\theta}_0$, the estimator is MS-unbiased. As a consequence, the bias $\bm{b}$ is zero and the MMSE and error covariance $C_p(\hat{\bm{\theta}}, \bm{\theta}_0)$ of the estimator $\hat{\bm{\theta}}$ coincide.}

\subsection{Empirical Validation of Model Usefulness and Bounds}
{An important observation is that equations \eqref{eq_MCRB} and \eqref{eq_bias} can be employed both to validate the computed MCRB and to verify the MS-unbiasedness of the implemented MMLE. Additionally, \eqref{eq_difference} is useful in corroborating the usefulness of the delay model in localization tasks through the lens of the MCRB. Note that the delay-based model and other similar models considering only few high amplitude scattering events are well-established and have been studied extensively in applications such as {direction of arrival} estimation, beamforming, and other fields involving the scattering of wave-like phenomena \cite{foldy1945multiple, lax1951multiple, gubernatis1977born, hudson1981use, semper2023misspecification}. As such, this verification through the MCRB is to be understood as another tool in the proverbial toolbox, with the important caveat that the true parameters of the true model must be known. This is only possible in simulations and during calibration.} 

{Under the assumption that the requisite regularity conditions hold, the KLD in \eqref{eq_kld} can minimized numerically in order to obtain the PTP $\bm{\theta}_0$, which can then be substituted into the Slepian formulae to compute the MCRB. The computed MCRB can then be compared against the \emph{Empirical Misspecified Mean Squared Error} (EMMSE) obtained from Monte Carlo simulations in the presence of noise. The EMMSE estimates the MMSE from \eqref{eq_bias} and is given by}
\begin{align} \label{eq_emmse}
	{{\text{EMMSE}(\hat{\bm{\theta}}, \bm{\theta}_0) = \frac{1}{N-1} \sum_{n=1}^N (\hat{\bm{\theta}}_n - \bm{\theta}_0)(\hat{\bm{\theta}}_n - \bm{\theta}_0)\trans},}
\end{align}
{based on $N$ estimates $\hat{\bm{\theta}}_n$ of the PTP $\bm{\theta}_0$. The empirical mean $\bar{\bm{\theta}}$ can be defined in a similar fashion. This EMMSE can then be decomposed as in \eqref{eq_bias} to corroborate that the bias with respect to the PTP $\bm{\theta}_0$ is negligible. This comparison simultaneously shows whether the chosen estimator is a MMLE and whether the EMMSE follows the MCRB.}

{An \emph{Empirical Mean Squared Error} (EMSE) can be defined analogously to \eqref{eq_emmse} and used in \eqref{eq_difference}. This provides an additional means of validating the computed MCRB and chosen estimator: if the estimator is a MMLE and the MCRB is computed correctly, equality is achieved in \eqref{eq_difference} in the limit as the number of realizations goes to infinity. As an added benefit, the computation of the difference vector $\bm{r}$ offers insight into the usefulness of the chosen misspecified model. If the entries of $\bm{r}$ are small, the delay model correctly conveys scatterer location information in spite of ignoring beam spread and multiple scattering. After this twofold validation, the MCRB can be used in further numerical simulations in which noise is accounted for through the noise covariance $\bm{\Sigma}$ instead of Monte Carlo trials, reducing the computational load.} 

%% file: sections/simulations.tex
\section{Simulations} \label{section_simulations}
In the numerical simulations presented next, inspiration is taken from ultrasound NDT. In particular, the simulations concern themselves with the task of locating \emph{defects} at which the \emph{Speed of Sound} (SoS) deviates from a known background value. The background medium is set to have an SoS of $c_0 = \SI{6400}{\meter\per\second}$ and no attenuation, i.e. $\beta_0 = 0$. A total of $U = 2$ point scatterers are placed in the medium, meaning the frequency flat contrast $\gamma$ is nonzero at exactly two locations $\bm{x}_u$ as discussed previously. Both defects are set to have the same wavenumber $k$. 

For the excitation $s(\omega, \bm{x}) = a(\omega)\delta(\bm{x} - \bm{x}_t)$ in \eqref{eq_hh} and $P(\omega)$ in \eqref{eq_simple}, a modulated Gaussian of the form
\begin{align}
	a(\omega) = P(\omega) = \frac{f_s}{2} \sqrt{ \frac{\pi}{\alpha} } \exp \left( \frac{-\pi^2}{\alpha} (f - f_c)^2 + \jmath \phi \right)
	\label{eq_gabor}
\end{align} 
is chosen. The parameters for the pulse shape described in \eqref{eq_gabor} are chosen so that they correspond to a reference real world measurement. The sampling frequency is set to $f_s = \SI{40}{\mega\hertz}$, the bandwidth is controlled by the factor $\alpha = (\SI{4.67}{\mega\hertz})^2$, the carrier frequency $f_c$ is $\SI{4.55}{\mega\hertz}$, and the phase offset $\phi$ has a value of $\SI{-2.61}{\radian}$. Although the simulations are carried out in the frequency domain, a reference number of time domain samples $N_t = 601$ is used so that the frequency resolution is given by $\Delta_\omega = \SI[parse-numbers=false]{f_s/N_t}{\hertz}$. In the frequency domain, $N_\omega = 161$ frequency bins in the bandwidth $[\SI{0.25}{\mega\hertz}, \SI{10.65}{\mega\hertz}]$ are considered. When noise is present, zero mean, circularly symmetric, additive white Gaussian noise is employed, meaning the noise covariance is of the form $\bm{\Sigma} = \sigma^2 \bm{I}_{N_\omega \cdot M_R \cdot M_T}$, with $\sigma^2 = 3$.

\begin{figure}
	\centering
	\includegraphics[width=1\columnwidth, trim = {0cm 0cm 1cm 0.5cm}, clip]{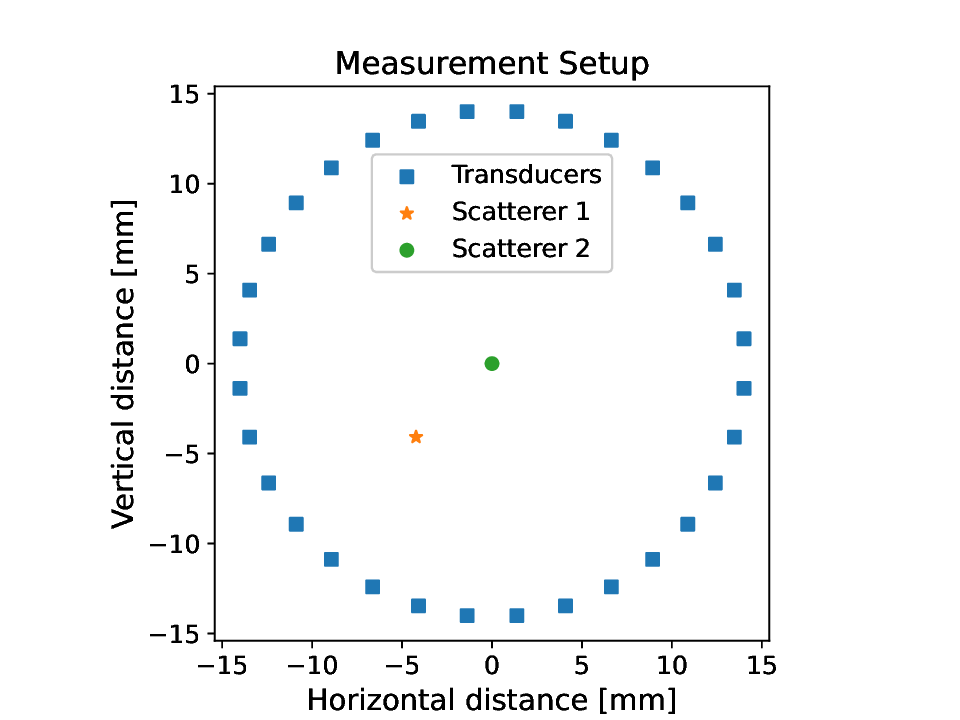}
	\caption{Measurement setup for the numerical simulations. A circular array surrounds a region of interest with two scatterers. The scatterer represented with a star is moved to different positions, while the circular one stays fixed. Transducers are shown as blue squares.}
	\label{fig_setup}
	\vspace{-0.5cm}
\end{figure}
 
The geometric parameters are chosen in terms of the reference wavelength $\lambda_0 = c_0/f_c$ as follows. The scatterers are located within a ROI of size $N_z \times N_x = 81 \times 81$ extending over a square region of size $(10 \lambda_0)^2$ with resolution $\Delta_z = \Delta_x = \lambda_0/8$. A total of 32 transducers are placed around the ROI along a circle of radius $10 \lambda_0$ at regular intervals. An illustration of the geometry is shown in \Cref{fig_setup}. For the sake of simplicity, the transducers do not interact with the medium and therefore produce no further reflections.

The simulation scenarios are constructed by varying the parameters as follows. One defect is moved to each of the possible $81 \times 81$ positions. The second scatterer is kept fixed {at} the origin, i.e. $\bm{x}_2 = [0, 0]\trans$. For each of the resulting configurations, the MCRB is computed. In this computation, {the true parameters are given by $\bm{\phi} = [\bm{x}_1\trans, \bm{x}_2\trans, \vert \gamma_1 \vert, \vert \gamma_2 \vert, \arg(\gamma_1), \arg(\gamma_2)]\trans$, while the parameters of interest are $\bm{\theta} = [ \bm{x}\trans_1, \bm{x}\trans_2, a_1, a_2, \theta_1, \theta_2 ]\trans$}, where $q_u = a_u \exp(\jmath \theta_u)$ is the scattering coefficient of each scatterer and $\bm{x}_u = [x_u ,\, z_u]\trans$ are the scatterer locations. This means that $\text{MCRB}(\bm{\theta}_0)$ is of size ${8 \times 8}$.

In this work, our interest lies in defect localization. In light of this, the quantity
\begin{align} \label{eq_Q}
	{Q = \sum_{i=1}^4 \sqrt{\bm{Q}_{ii}}}
\end{align}
{is taken as a localization performance indicator. The $\bm{Q}$ matrix is to be substituted by the classical CRB or any of the quantities involved in \eqref{eq_MCRB}, \eqref{eq_difference}, or \eqref{eq_bias} (e.g. $\text{MCRB}(\bm{\theta}_0)$ or $\bm{rr}\trans$). This means that $\bm{Q}$ is real-valued and is of size $8 \times 8$ if it is defined over $\bm{\theta}$ or $\bm{\phi}$, or of size $4 \times 4$ if the projection $(\cdot)^s$ was employed as is the case with $\bm{rr}\trans$. Furthermore, $\bm{Q}_{ii}$ with $i \in \{1, 2, 3, 4\}$ refers to the first four diagonal entries of the $\bm{Q}$ matrix, which in all cases corresponds to the location parameters $x_1$, $z_1$, $x_2$, $z_2$ of the two scatterers.}

{With this definition, the scalar $Q$ represents a form of absolute error. Specifically, depending on the chosen matrix $\bm{Q}$, the $Q$-value represents a sum of standard deviations, a sum of absolute differences, or a sum of absolute biases, and is equivalent to computing elementwise square roots followed by the traces of \eqref{eq_MCRB}, \eqref{eq_difference}, and \eqref{eq_bias} over the subspace $(\cdot)^s$ of location parameters. This enables the creation of an image of $81 \times 81$ pixels, {referred to as a $Q$-image,} where the location of each pixel corresponds to the coordinates $\bm{x}_1$ of the moving scatterer as given in $\bm{\phi}$ and the pixel intensity is given by the corresponding $Q$-value which has units $[\unit{\micro\meter}]$. The individual $Q$-values will be referred to by the quantity they represent, e.g. ``Bias'' meaning that $\bm{Q} = \bm{bb}\trans$ is substituted into \eqref{eq_Q}.}

\subsection{Computational Details} \label{subsec_comp}
When generating fields according to the Foldy-Lax model, equations \eqref{eq_ftot} and \eqref{eq_gamma} are treated differently. If $\Groi$ in \eqref{eq_ftot} is computed simply by taking samples of the Green's function, the condition number is poor and inversion, including direct inversion, is challenging. Instead, the matrix is constructed following the 2D trapezoid rule. This can be interpreted either as a smoothing of the Green's function, or as letting each defect consist of a group of four closely spaced defects. As no inversion is involved in \eqref{eq_gamma}, $\Garray$ is constructed from samples of the Green's function.

{Both the {PTP} $\bm{\theta}_0$ and estimates $\hat{\bm{\theta}}$ are required for the computation of the MCRB. Due to the noise statistics, the minimization of the KLD \eqref{eq_kld}, which yields the PTP $\bm{\theta}_0$, takes the nonlinear least squares form}
\begin{align} \label{eq_kldmin}
	\bm{\theta}_0 = \argmin_{\bm{\theta}} \Vert \bm{s} - \bm{\mu}(\bm{\theta}) \Vert_2^2.
\end{align}
{Asymptotically MS-unbiased estimates can be obtained by minimizing the negative of the MLL \eqref{eq_MLL}, which is equivalent to}
\begin{align}
	\hat{\bm{\theta}}_\text{MML} = \argmin_{\bm{\theta}} \Vert \bm{y} - \bm{\mu}(\bm{\theta}) \Vert_2^2.
	\label{eq_nlls}
\end{align}
A two step procedure is adopted in solving for $\bm{\theta}_0$ and $\hat{\bm{\theta}}_\text{MML}$. Note, however, that the procedure described next is not suitable for reconstruction and localization in practice, but is only employed given the amount of prior information available when generating simulations.

\begin{figure*}
	\centering
	\begin{subfigure}{0.49\textwidth}
		\includegraphics[width=1\columnwidth, trim = {0.2cm 0.2cm 0cm 0.1cm}, clip]{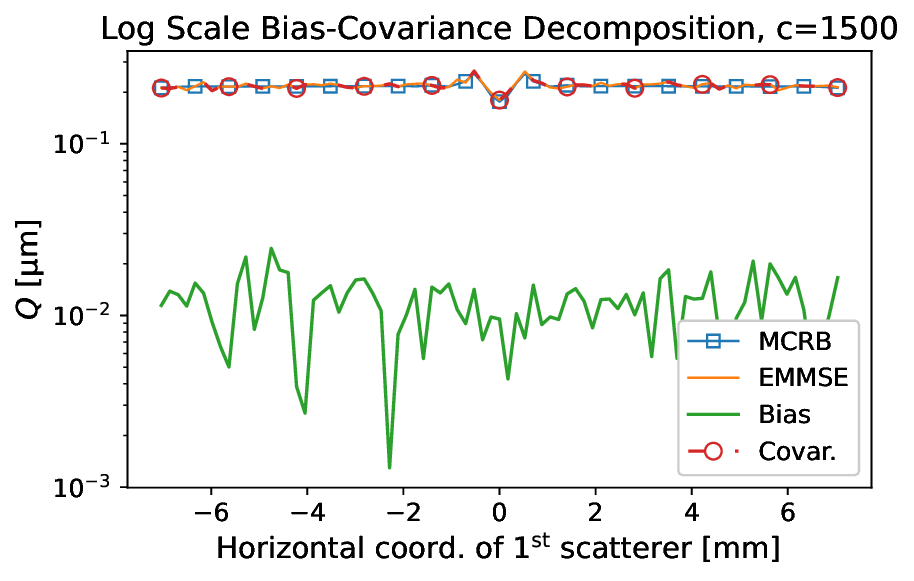}
		\caption{}
		\label{fig_EMMSE1500_decomp}
	\end{subfigure}%
	\begin{subfigure}{0.49\textwidth}
		\includegraphics[width=1\columnwidth, trim = {0.2cm 0.2cm 0cm 0.1cm}, clip]{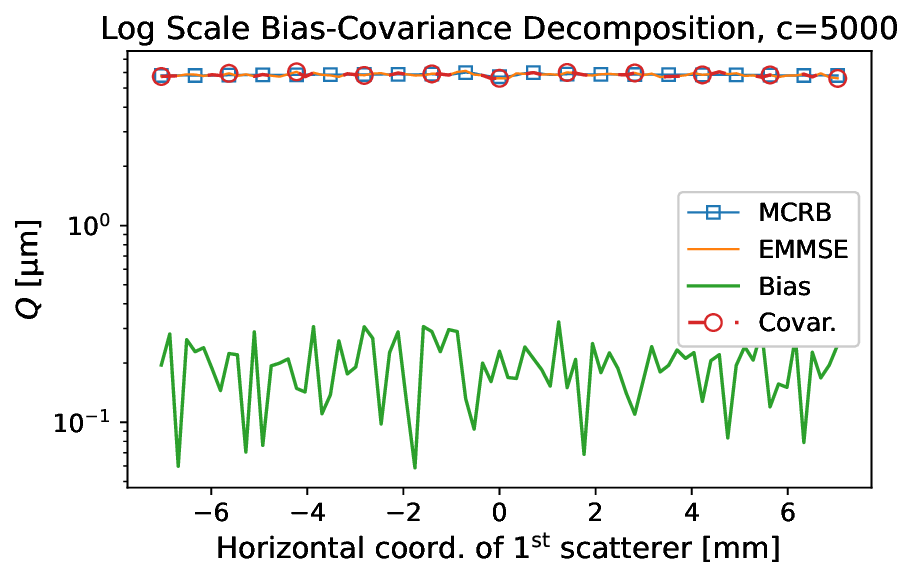}
		\caption{}
		\label{fig_EMMSE5000_decomp}
	\end{subfigure}
	
	\begin{subfigure}{0.49\textwidth}
		\includegraphics[width=1\columnwidth, trim = {0cm 1cm 1cm 1.3cm}, clip]{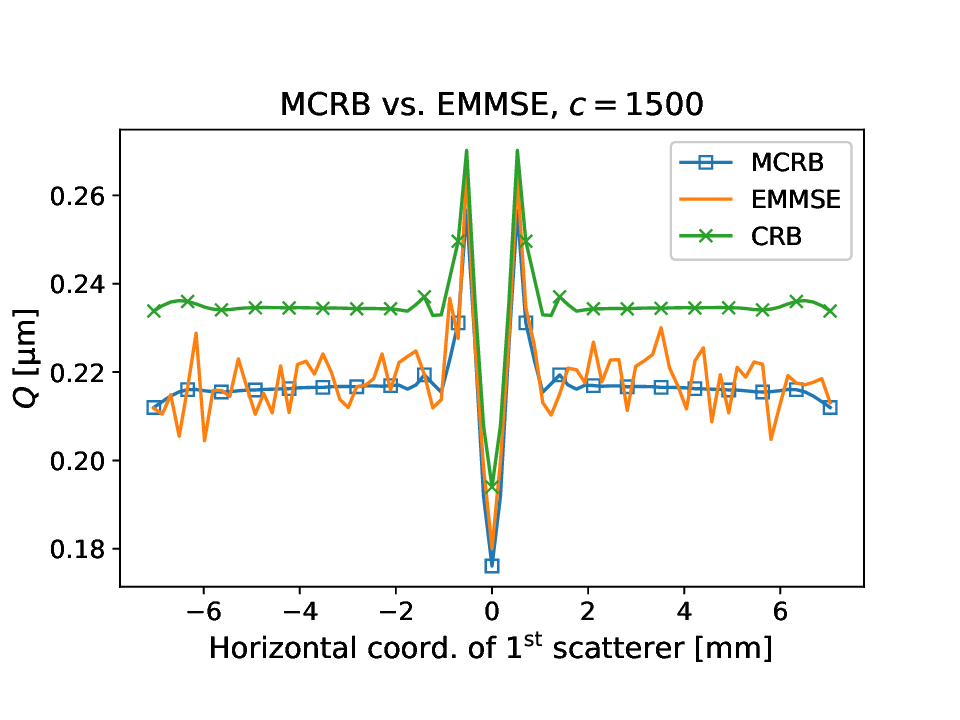}
		\caption{}
		\label{fig_EMMSE1500}
	\end{subfigure}%
	\begin{subfigure}{0.49\textwidth}
		\includegraphics[width=1\columnwidth, trim = {0cm 1cm 1cm 1.3cm}, clip]{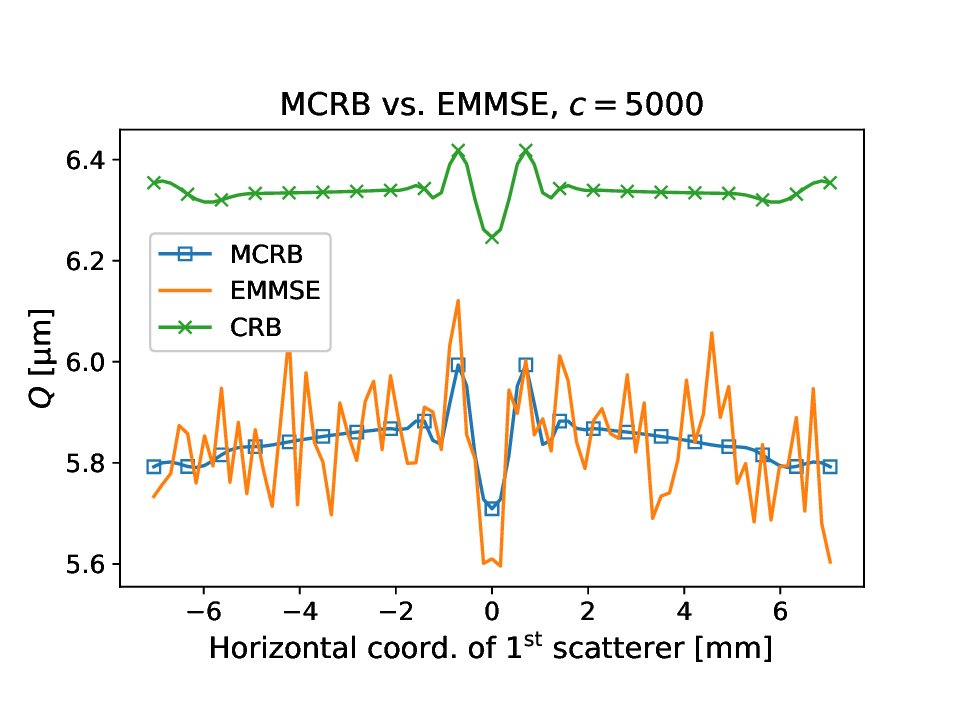}
		\caption{}
		\label{fig_EMMSE5000}
	\end{subfigure}
	\caption{{Comparison between EMMSE and MCRB for different simulation parameters. The EMMSE is computed for 200 noise realizations when $\boldsymbol{c} = \mathbf{\qty{1500}{\meter \per \second}}$ and 500 realizations for $\boldsymbol{c} = \mathbf{\SI{5000}{\meter \per \second}}$. \Cref{fig_EMMSE1500_decomp,fig_EMMSE5000_decomp} show the empirical bias-covariance decomposition of the EMMSE in logarithmic scale. The empirical bias is an order of magnitude smaller than the covariance. Since the EMMSE $\boldsymbol Q$-value is obtained through the Pythagorean addition of the empirical bias and covariance, the influence of the bias is negligible and the covariance coincides with the EMMSE. The EMMSE and covariance additionally coincide with the MCRB. In \Cref{fig_EMMSE1500,fig_EMMSE5000}, the agreement between EMMSE and MCRB is illustrated in linear scale. Even though more noise realizations are used in the computation of \Cref{fig_EMMSE5000}, the EMMSE appears noisier than in \Cref{fig_EMMSE1500} due to the lower SoS contrast and consequently lower SNR. The classical CRB is shown as a reference, and the EMMSE is observed to follow the MCRB instead of the CRB.}}
	\label{fig_EMMSE}
\end{figure*}

\blue{The pulse shape, the grid where the defects are located, and the number of defects are all available in simulations, allowing the construction of a linear model.} A dictionary matrix $\bm{A}\in \compl^{N_\omega \cdot M_R \cdot M_T \times 2}$ is built based on the discretized delay model given all the prior knowledge, letting each column correspond to the signal that would be observed if only a single scatterer were present. The only unknown quantities are the scattering amplitudes $q_u$, which can be obtained as
\begin{align}
 \bm{q}_{\text{LS}} = \argmin_{\bm{q}} \Vert \bm{y} - \bm{Aq} \Vert_2^2 = \bm{A}^\dagger \bm{y},
\end{align}
where $(\cdot)^\dagger$ denotes the pseudoinverse. The entries of $\bm{q}_\text{LS} \in \compl^{2}$ correspond to estimates of the scattering amplitudes $q_u$. When both the true and assumed model correspond to the delay model, this step suffices to obtain all the model parameters. However, the scatterer coordinates in the delay model in general do not coincide with those of the Helmholtz model. 

In the second step, the estimates are refined by using the \emph{Broyden-Fletcher-Goldfarb-Shanno} (BFGS) algorithm \cite{BFGS}. The target functions are the nonlinear least squares KLD and MML expressions in \eqref{eq_kldmin} and \eqref{eq_nlls}. {The true scatterer coordinates $\bm{x}_{1}$, $\bm{x}_{2}$ in $\bm{\phi}$, as well as $\bm{q}_\text{LS}$, are used as initial guesses for the parameters $\bm{\theta}$.} {We emphasize that this is possible in simulations, but the true parameters $\bm{\phi}$ of $\bm{s}$ are unknown in practice.} The derivatives involved in the BFGS algorithm are computed through automatic differentiation using the JAX library \cite{jax}.
  
\subsection{Validation}
Two scenarios are considered in the validation of the computed MCRBs and the related $Q$-images. In the first scenario, the defects are given an SoS $c = \SI{1500}{\meter \per \second}$, and the MMLE is computed following the aforementioned procedure. This is done for 200 noise realizations, after which the {EMMSE} of the estimates of the localization parameters is used to compute a cross section of a $Q$-image. Next, {the PTP is computed by minimizing the KLD.} Using \Cref{th_slepian}, the MCRB is computed based on the {PTP} and a noise covariance matrix $\bm{\Sigma} = \sigma^2 \bm{I}$, with $\sigma^2 = 3$. From this result, the corresponding $Q$-image cross section is computed. Finally, the {PTP} is taken as the input parameter of the delay model, which is now treated as the true model. This yields one last $Q$-image cross section corresponding to the classical CRB to be used as a reference. This procedure is then repeated for $c = \SI{5000}{\meter \per \second}$ for 500 noise realizations. The cross sections of the $Q$-images are constructed by allowing $x_1$ to vary and keeping $z_1 = \SI{-0.704}{\milli\meter}$ fixed. The cross sections are shown in \Cref{fig_EMMSE}.

{The {EMMSE}-based $Q$-values are observed to follow those of the MCRB, advocating for its validity. The empirical bias-covariance decomposition of the {EMMSE} in \Cref{fig_EMMSE1500_decomp,fig_EMMSE5000_decomp} illustrates that the empirical bias is negligible, meaning the estimator is practically MS-unbiased.} \Cref{fig_EMMSE1500} shows better agreement with the MCRB than \Cref{fig_EMMSE5000}, since the magnitude of the frequency flat contrast $\gamma$ increases as the difference between the background and defect SoS increases. This increases the scattering amplitude and \emph{Signal to Noise Ratio} (SNR) relative to the simulation scenarios with a lower SoS contrast. The {EMMSE} agrees with the MCRB instead of the classical CRB for both SoS values, as is to be expected in the presence of model mismatch. {Reiterating, \Cref{fig_EMMSE} is an application of \eqref{eq_MCRB} and \eqref{eq_bias} illustrating the validity of the computed MCRB and the MS-unbiasedness of the estimator.}

\begin{figure*}
	\centering
	\begin{subfigure}{0.49\textwidth}
		\includegraphics[width=1\columnwidth, trim = {0.2cm 0.2cm 0cm 0.1cm}, clip]{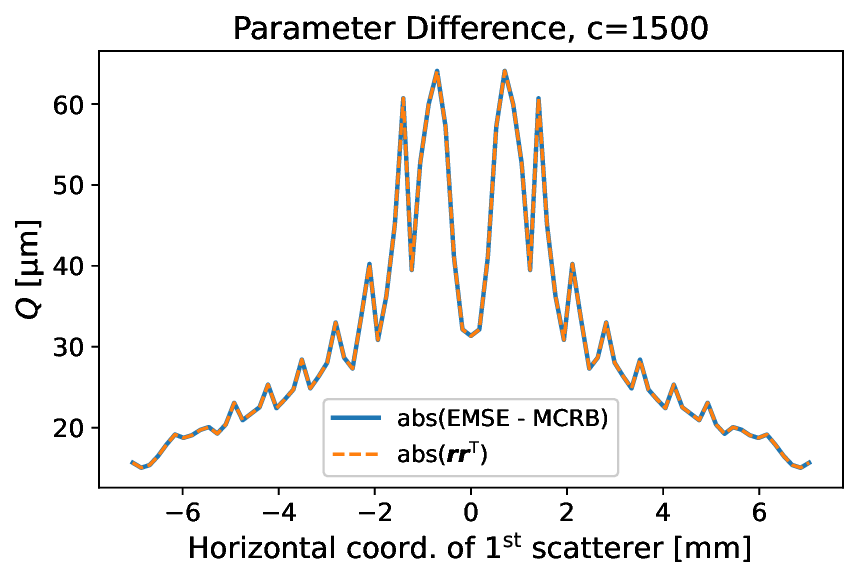}
		\caption{}
		\label{fig_difference1500_decomp}
	\end{subfigure}%
	\begin{subfigure}{0.49\textwidth}
		\includegraphics[width=1\columnwidth, trim = {0.2cm 0.2cm 0cm 0.1cm}, clip]{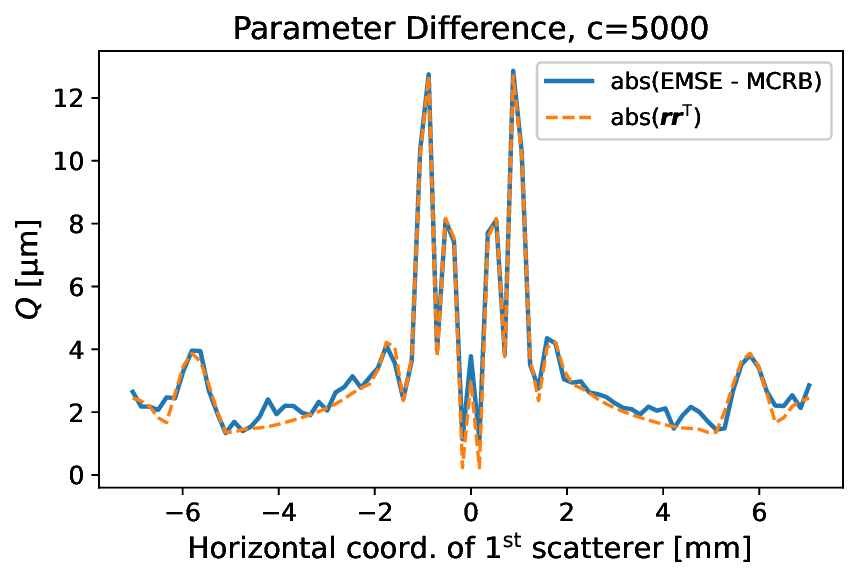}
		\caption{}
		\label{fig_difference5000_decomp}
	\end{subfigure}
	\caption{{Decomposition of the Empirical MSE based on \eqref{eq_difference}. Since the empirical bias is small as illustrated in \Cref{fig_EMMSE}, agreement between the curves means that the error covariance coincides with the MCRB. The small $\boldsymbol{Q}$-values compared to the true parameter values illustrates that the PTS $\boldsymbol{\bm{\theta}}_{\boldsymbol{0}}$ of the model ignoring multiple scattering still provides valuable localization information.}}
	\label{fig_difference}
\end{figure*}

{Using the same parameters as in \Cref{fig_EMMSE}, the EMSE and empirical parameter error $\bm{r}$ over the subspace of defect location parameters are computed next. \Cref{fig_difference} shows a good match between $\text{abs}(\bm{rr}\trans)$ and $\text{abs}(\text{MSE} - \text{MCRB})$, which is possible for MS-unbiased estimators whose error covariance coincides with the MCRB. This corroborates the findings illustrated in \Cref{fig_EMMSE}. Additionally, the $Q$-values are noticeably small. Consider the worst case scenario in which the parameter error plotted in \Cref{fig_difference} is attributed to a single location parameter out of the total four, e.g. all of the error is due to $x_1$. Even in this case, the parameter error is two to three orders of magnitude smaller than the value of the parameter, and is therefore negligible except for very closely spaced scatterers. This serves as an alternative corroboration of the widespread success of the delay-based model. As a final remark regarding validation, notice that the location parameter difference in \Cref{fig_difference5000_decomp} is smaller than that in \Cref{fig_difference1500_decomp}. This coincides with the notion that the degree of misspecification of multiple scattering is lower when the SoS contrast is smaller, otherwise referred to as the weak scattering regime.}

The remaining simulations are carried out in the absence of noise, recalling that the MCRB depends on the noise only through the covariance matrix $\bm{\Sigma}$ which can be introduced ex post facto. This allows the construction of the full $Q$-images for all SoS values.

\begin{figure*}
	\centering
	\includegraphics[width=1\textwidth, trim = {8.6cm 6.2cm 8cm 4.5cm}, clip]{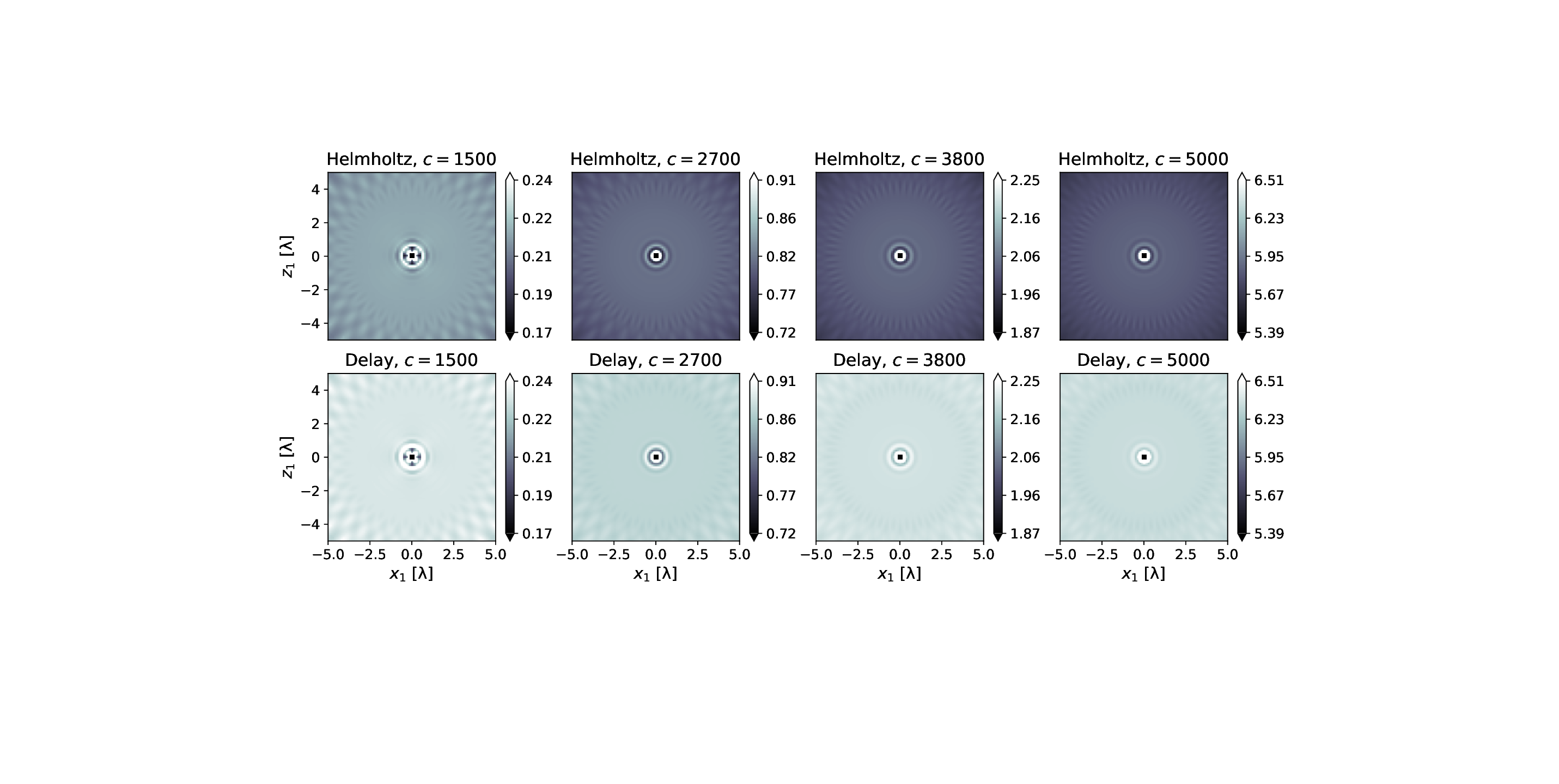}
	\caption{Illustration of the $\boldsymbol{Q}$-images for the MCRB (top row) and CRB (bottom row) for 8 different simulation scenarios. Each image corresponds to a fixed true model (Helmholtz or delay) and a defect speed of sound $\boldsymbol{c}$ with units $\mathbf{{\SI[parse-numbers=false]{}{\meter\per\second}}}$. The assumed model during estimation is the delay model in all cases. The color bar is shown in $\mathbf{{\unit{\micro\meter}}}$. The values of the MCRB $\boldsymbol{Q}$-images are lower than those of the classical CRB. The value of the center pixel is undefined, as the two scatterers are at the exact same location.}
	\label{fig_allMCRBs}
\end{figure*}

\subsection{Simulation Results}
Having validated the numerical computation of the MCRB, we move on to simulations without Monte Carlo noise realizations. We consider the following scenarios. The defects are allowed to have wavenumbers defined by $c \in \{ \SI{1500}{\meter\per\second}, \SI{2700}{\meter\per\second}, \SI{3800}{\meter\per\second}, \SI{5000}{\meter\per\second} \}$, $\beta = 0$. {Recalling that the background SoS is $c_0 = \SI{6400}{\meter \per \second}$}, scenarios ranging from high to low SoS contrasts, and therefore wavenumber contrasts, are considered. As the wavenumber contrast decreases, the magnitude of the scattered field and the effect of multiple scattering decreases, the three models discussed in \Cref{section_models} become more similar, and the degree of misspecification decreases. In order to study this behavior, {two true models are considered}: the Helmholtz model and the delay model. 

{As was done during the validation, the PTP is taken as the input parameter of the delay model.} {More details on this choice are presented in} \Cref{section_discussion}. This yields a total of 8 different simulation scenarios: for each of the two true models (the Helmholtz and delay model), each of the four defect speeds of sound is considered and a corresponding $Q$-image is computed.

{To reiterate, the simulation scenarios consist of all possible combinations of four defect SoS values $c \in \{ \SI{1500}{\meter\per\second}, \SI{2700}{\meter\per\second}, \SI{3800}{\meter\per\second}, \SI{5000}{\meter\per\second} \}$ paired with the delay and Helmholtz models, yielding a total of eight scenarios. In all scenarios, the assumed model during estimation is the delay model, meaning that there is misspecification when the true model is the Helmholtz model}. The $Q$-images for the eight simulation scenarios involving different models and scatterer parameters are presented in \Cref{fig_allMCRBs}. The axes are shown in wavelengths, i.e. $L/\lambda_0$ for a length $L$, so as to facilitate the interpretation of the distance between the scatterers. 

{For a fixed SoS $c$, t}he $Q$-images appear to have an overall similar appearance {regardless of the true model}. {However, t}he MCRB values are overall lower than those of the CRB. This means that, in the simulation scenarios, the presence of multiple scattering in the data results in better {estimation} performance {in terms of the error covariance of $\hat{\bm{\theta}}$, across all values of $c$ and $\bm{\phi}$,} than would be predicted by the classical CRB. This observation is to be interpreted carefully, and is discussed in detail next.

%% file: sections/discussion.tex
\section{Discussion} \label{section_discussion}
The simulation scenarios in this work hint at an overall performance improvement in localization accuracy in the presence of multiple scattering, even when the wavenumber contrast is low. At first glance, this appears to contradict findings in the literature which state that the presence of multiple scattering may be beneficial or detrimental depending on the configuration of the sensors and scatterers \cite{crb_mono1, crb_mono2, crb_resolved, crb_optics, born_crb}. However, these works consider single transmitter and receiver scenario with harmonic excitation, motivated by the search of analytic expressions. In contrast, our simulations employ a sensor array and a realistic signal with a broad bandwidth, and we study the impact of model mismatch directly. Although we don't present analytic results, the overall procedure for the computation of numerical results and their interpretation can be applied to any scenario.

More importantly, {when mismatch is considered in these works, it is studied through the MSE. This automatically introduces a parameter difference $\bm{r}$ that, in these works, is referred to as a bias.} {As discussed in} \Cref{subsec_missp}, {this difference should not be interpreted as a bias in the context of misspecification. Instead, the presence of misspecification turns the MLE into the MMLE. In the context of misspecification and the MCRB, the results reported in \cite{crb_mono1, crb_mono2, crb_resolved, crb_optics, born_crb} contain a combination of parameter space differences and bias with respect to the PTP. The interpretation of the findings in the present work and a comparison to the aforementioned prior works constitutes the remainder of this discussion.} 

{Before jumping into the discussion, however, the question arises as to why the CRB is consistently higher than the MCRB. Although additional factors may be involved, the connection between total signal energy and multiple scattering can be highlighted. In the present investigation on the misspecification of multiple scattering, model mismatch directly affects the SNR.} As was illustrated in \Cref{fig_model_comparison}, the {delay model often matches the first scattering event in the Helmholtz model}. However, further scattering events are ignored, meaning that the energy content of {signals following these models is different}. Normalization and scaling are possible, {but this difference in signal energy is inherent to the model choice. The presence of multiple scattering can thus increase the SNR when distinct echoes do not overlap or when they interfere constructively with each other. With this observation out of the way, we first address the matter of parameter differences and bias}.

\subsection{Bias or Parameter Difference?}
Moving onto the interpretation of the results, we reiterate the accuracy of the MCRB when describing the {MMSE}, {and not the MSE. This observation is crucial because, in the presence of misspecification, the MSE involves the computation of differences on intersecting subspaces of the parameter spaces of distinct models. This motivates the usage of the projection $(\cdot)^s$ onto the subspace of location parameters when considering the MSE throughout this work.} Instead, the MCRB as defined in \Cref{def_MCRB} incorporates misspecification by considering the error covariance with respect to the {PTP}, i.e. within a single parameter space.

{In the delay model, only the scatterer location parameters overlap with the parameter space of the true model. This means that, for a true model $\bm{s}_\text{Helmholtz}$ (where the subscript has be added for clarity) with true parameter $\bm{\phi}$, the quantity $\bm{\mu}_{\text{delay}}(\bm{\phi})$ has no useful interpretation. In contrast, the aforementioned works use the Born approximation, which shares its parameter space with the Helmholtz model. It may then appear more reasonable to evaluate the Born model at the true parameter $\bm{\phi}$, but the same property still holds: even though $\bm{\mu}_{\text{Born}}(\bm{\phi})$ can be properly evaluated and appears similar to $\bm{s}_\text{Helmholtz}(\bm{\phi})$, the two are not directly related since the models are distinct.} 

{This can be observed e.g. in \cite{born_crb}, where it was noted that the usage of the Born model during estimation introduces a ``bias''. The observed bias was computed based on the MSE and therefore actually refers to the parameter difference $\bm{r}$, which in their case is properly defined over the entire parameter space without needing the projection $(\cdot)^s$. The observed parameter difference means that, although the models share the same parameter space, the presence of misspecification in the model has affected the estimation procedure. As mentioned previously, this can be interpreted as misspecification ``turning the MLE into the MMLE'', which we make precise as follows.}

{Recall that the Mismatched Maximum Likelihood Estimator under misspecification of multiple scattering was formulated in \eqref{eq_nlls} as
\begin{align*}
	\hat{\bm{\theta}}_{\text{MML}} = \argmin_{\bm{\theta}} \Vert \bm{y} - \bm{\mu}(\bm{\theta}) \Vert_2^2,
\end{align*}
with $\bm{y} = \bm{s}(\bm{\phi}) + \bm{n} = \bm{\psi}_\text{s}(\bm{\phi}) + \bm{n}$ and $\bm{\mu}(\bm{\theta}) = \bm{\psi}_\text{D}(\bm{\theta})$. The aforementioned quantities can be explicitly substituted into the minimization problem so that
\begin{align} \label{eq_mmle_example}
	\hat{\bm{\theta}}_{\text{MML}} =
	\argmin_{\bm{\theta}} \Vert \bm{\psi}_\text{s}(\bm{\phi}) + \bm{n} - \bm{\psi}_\text{D}(\bm{\theta}) \Vert_2^2.
\end{align}
This estimator is asymptotically MS-unbiased with expected value $\bm{\theta}_0$. What happens if we now change the model $\bm{s}$ to the delay model without modifying the noise statistics? Doing this yields the expression
\begin{align} \label{eq_mle_example}
	\hat{\bm{\theta}}_{\text{ML}} =
	\argmin_{\bm{\theta}} \Vert \bm{\psi}_\text{D}(\bm{\theta}^*) + \bm{n} - \bm{\psi}_\text{D}(\bm{\theta}) \Vert_2^2,
\end{align}
in which $\bm{\theta}^*$ is the true parameter of the new true model $\bm{\psi}_\text{D}$. This expression clearly has no misspecification and describes the Maximum Likelihood Estimator.} 

{When only the true mean of the data is changed, the MMLE turns into the MLE. As such, we can say that ``In the presence of misspecification, since the MLE turns into the MMLE (in the sense described previously), the estimation task automatically deals with the search of the PTP $\bm{\theta}_0$, and not the search of the true parameter $\bm{\phi}$ of the true model. In ultrasound localization, the presence of model misspecification means that the estimated locations and scattering amplitudes generally do not match the ground truth even if the misspecified model appears to have the same parameter space as the true one.'' This directly explains the observed ``bias'', which as mentioned previously is technically the parameter difference $\bm{r}$ introduced by the misspecification. We can then add that ``The magnitude of the resulting parameter difference $\bm{r}$ (when it is defined) determines whether the chosen misspecified model is useful for the estimation task. In the context of ultrasound localization, a small parameter difference means defects can be located accurately despite the model mismatch.'' We follow up on this point by studying additional properties of the MLE and MMLE.}

\subsection{Using the PTP as a Reference}
{Next, the asymptotic behavior of the MLE and MMLE can be studied to justify the usage of the PTP when comparing a misspecified case against a correctly specified one. Referring to the PTP as $\bm{\theta}_0$, the true parameter of the Helmholtz model as $\bm{\phi}$, and completely separately considering a correctly specified case where the true model is the delay model with true parameter $\bm{\theta}^*$ as was done in \eqref{eq_mmle_example} and \eqref{eq_mle_example}, the following properties hold. When the MMLE is employed on $M$ noise realizations,
\begin{align*}
	\hat{\bm{\theta}}_{\text{MML}} - \bm{\theta}_0 \underset{M \to \infty}{\overset{d}{\to}} \mathcal{N}(\bm{0}, \text{MCRB}(\bm{\theta}_0)),
\end{align*}
where $\underset{M \to \infty}{\overset{d}{\to}}$ denotes convergence in distribution \cite{MCRBreal, MCRB}. Similarly, the MLE has the property that
\begin{align*}
	\hat{\bm{\theta}}_{\text{ML}} - \bm{\theta}^* \underset{M \to \infty}{\overset{d}{\to}} \mathcal{N}(\bm{0}, \text{CRB}(\bm{\theta}^*)).
\end{align*}
These relationships elucidate that both estimators (asymptotically) belong to the same distribution family, but have different parameters.}

{If we now let $\bm{\theta}^* = \bm{\theta}_0$, we observe that the MLE and MMLE have the same mean, but different variances. Correspondingly, choosing to evaluate the delay model at the PTP means that, in this new and correctly specified scenario, the MLE will (in expectation) produce the same parameter estimate that the MMLE produces in the presence of misspecification. As a consequence, the two estimators differ only in their covariance, allowing for a straightforward comparison between the misspecified and correctly specified cases. This justifies our usage of the PTP in the delay model as was done in \Cref{fig_EMMSE} and \Cref{fig_allMCRBs}.}

{In contrast to this, if one were to evaluate the correctly specified case at a different parameter, say at $\bm{\phi$} as was done in the works \cite{crb_mono1, crb_mono2, crb_resolved, crb_optics, born_crb}, the MMLE and MLE differ both in mean and covariance and a direct comparison would require the computation of a proper distance or divergence between their respective distributions. We can say that ``A reference scenario can be constructed by treating the misspecified model as if it were the true model. Evaluating this new scenario at the PTP isolates the effect of misspecification to the covariance of the MLE and MMLE, making a comparison between the two straightforward. In ultrasound localization, this means that the location parameters and scattering amplitudes in both scenarios will have the same mean values, but different covariance due to the misspecification. Due to the previous discussion point, neither of these estimators will generally coincide with the ground truth.''} 

{Choosing to create a reference scenario based on the assumed, misspecified model evaluated at the PTP has additional consequences which we discuss next.}

\subsection{Behavior of the CRB}
{As a final point, we once again draw attention to \Cref{fig_EMMSE,fig_allMCRBs}. In these figures, the MCRB appears to be close to a vertical shift of the CRB, which begs the question of whether the computational cost of the MCRB is warranted. {This phenomenon is directly related to the previous point, where it was stated that the reference scenario was constructed by treating the delay model as a new, correctly specified model and evaluating it at the PTP. Doing so eliminates the misspecification and allows the computation of the classical CRB}. 

As shown in \Cref{def_MCRB,def_MLL}, the MCRB depends explicitly on the \blue{PTP} $\bm{\theta}_0$, which itself depends on the misspecified and true models $f_X(\bm{x} \vert \bm{\theta})$ and $p_X(\bm{x})$. Recalling \Cref{th_MMLE} and equations \eqref{eq_kld} and \eqref{eq_MLL}, we highlight that the \blue{PTP} is obtained by projecting the true model onto the misspecified one. {This becomes apparent when observing the nonlinear least squares forms in \eqref{eq_kldmin} and \eqref{eq_nlls}. In this manner, both the choice of true model and its true parameters will affect the PTP $\bm{\theta}_0$.} 

As a result of computing the CRB based on a reference scenario evaluated at PTP, the behavior of the CRB is dictated not only by the assumed model, but also by the true model evaluated at the true parameter. To illustrate this, we provide one final example in \Cref{fig_b_vs_h} in which a third scenario has been considered. The Born model is now taken as the true model, while the delay model remains as the assumed one. {The {PTP} is computed once again using the same ground truth parameters $\bm{\phi}$ as for the Helmholtz model in \Cref{fig_EMMSE1500}. The Helmholtz and Born models superficially appear to have the same parameter space. The usage of a different model true model, however, introduces a parameter space difference and yields a different PTP.}

\begin{figure}
	\centering
	\includegraphics[width=1\columnwidth, trim = {0cm 0cm 0cm 0cm}, clip]{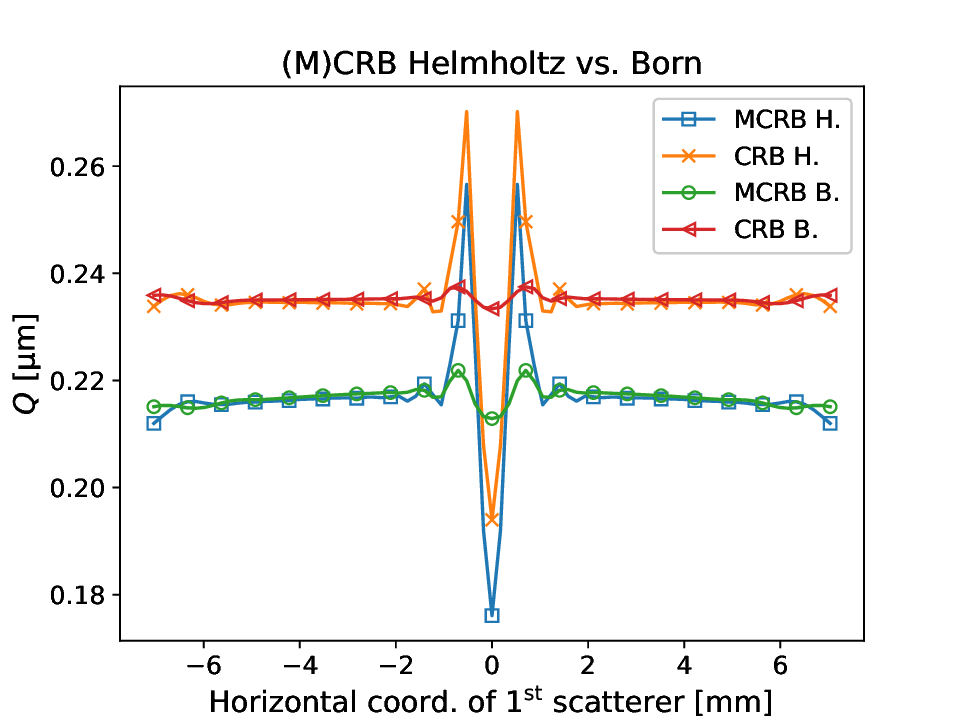}
	\caption{Illustration of the MCRB and CRB when the true model is the Helmholtz (H.) or Born (B.) model. The classical CRB, where the true model is the delay model, is shown to behave similarly to MCRB B. and MCRB H. simply by changing its parameters to the corresponding PTP.}
	\label{fig_b_vs_h}
\end{figure}

{We observe that the classical CRB for the delay model can behave either as the MCRB using the Helmholtz model or the MCRB using the Born model, depending on which of these is used as the true model from which the PTP is obtained.} \blue{As such, it is necessary to generate data based on the Helmholtz model to obtain ``realistic'' CRB values in the first place, though they are nevertheless inapplicable due to the misspecification. In particular, if the true model is the Born model, the resulting PTP and classical CRB erroneously convey the message that the estimation task remains simple even when the defects are closely spaced. This is not the case when the true model contains multiple scattering, as shown by the curves labelled ``MCRB H.'' and ``CRB H.'' in \Cref{fig_b_vs_h}.} 

Note that once the data has been generated based on the Helmholtz model and the MMLE has been employed, the computational cost of the CRB and the MCRB is comparable. It is therefore preferable to simply use the MCRB instead. Importantly, experimental design relies on accurately predicting the achievable performance in challenging scenarios, and so the MCRB should be employed over the CRB when misspecification is present. 

{This can be summarized lightheartedly as ``{Constructing reference scenarios to evaluate the impact of model mismatch is a treacherous task in which seemingly innocuous choices regarding the parameters can make a direct comparison of models and bounds difficult or impossible.} Parameter differences due to model mismatch and the statistical properties of the MLE and MMLE must be considered. In a poorly constructed reference scenario, neither the expected value nor the covariance of the corresponding MLE task convey any information about the misspecified estimation task. Constructing a reference scenario by evaluating the misspecified model at the PTP is a good choice since, in addition to granting the MLE and MMLE the same expected value, it makes their covariance behave similarly. Nevertheless, in the presence of misspecification, the CRB is not a valid lower bound for any parameter $\bm{\theta}$ of the assumed model. A proper evaluation of the impact of multiple scattering on ultrasound localization tasks should consider all of these factors.''}

%% file: sections/conclusion.tex
\section{Conclusion}
In this work, we have employed the Misspecified CRB to study the impact of the misspecification of multiple scattering in localization tasks where the measurement data stems from a scalar wave phenomenon. We have employed the generalized Slepian formulae to study simulated ultrasound tomography data with the goal of contributing to the discussion on whether multiple scattering is beneficial or detrimental in localization tasks. It was observed that the achievable scatterer localization performance \blue{under misspecification of multiple scattering} is better than would be predicted by the classical {CRB}. {However, this statement is not to be taken lightly.}

{The discussion presented in this work aims to highlight the difficulty and nuance involved in a proper evaluation of the impact of multiple scattering in localization tasks. The construction of useful reference scenarios and comparisons is especially daunting, since the influence of parameter space differences and the statistical properties of the MLE and MMLE must be considered simultaneously. To exemplify this, we have recontextualized earlier studies on the impact of multiple scattering through the lens of the MCRB. In doing so, it becomes clear that errors previously attributed to bias are instead caused by parameter space differences between mismatched models. Through this distinction, the MCRB is able to better isolate and describe the influence of misspecification on the localization task than the classical CRB. We advocate for the usage of the PTP in the construction of reference scenarios without model mismatch, as it endows the MLE with desirable statistical properties that account for this parameter space difference.}


